\documentclass{emulateapj}
\usepackage{apjfonts}
\usepackage{natbib}
\usepackage{graphicx}
\usepackage{epstopdf}
\usepackage{multirow} 
\usepackage{longtable,lscape} 
\usepackage{color}

\newcommand{\M}{$\log M_\star/M_\odot$}

\shorttitle{The mass function predicted by simulations}
\shortauthors{Vulcani et al.}

\begin{document}
\title{What do simulations predict for the galaxy stellar mass function and its evolution in different environments?} 

\author{Benedetta Vulcani\altaffilmark{1}}
\author{Gabriella De Lucia\altaffilmark{2}}
\author{Bianca M. Poggianti\altaffilmark{3}}
\author{Kevin Bundy\altaffilmark{1}}
\author{Surhud More\altaffilmark{1}}
\author{Rosa Calvi\altaffilmark{4}}

\affil{Kavli Institute for the Physics and Mathematics of the Universe (WPI), Todai Institutes for Advanced Study, the University of Tokyo, Kashiwa, 277-8582, Japan}
\affil{INAF - Astronomical Observatory of Trieste, 34143 Trieste, Italy}
\affil{INAF - Astronomical Observatory of Padova, 35122 Padova, Italy}
\affil{Astronomical Department, Padova University, 35122 Padova, Italy}

\email{benedetta.vulcani@impu.jp}

\begin{abstract}
We present a comparison between the observed  galaxy stellar mass function and the one predicted from the \cite{deluciablaiz07} semi-analytic model applied to the Millennium Simulation, for cluster satellites and galaxies in the field (meant as a wide portion of the sky, including all environments), in the local universe ($z\sim0.06$) and at intermediate redshift ($z\sim0.6$),  with the aim to shed light on the processes which regulate the mass distribution in different environments. 
While the mass functions in the field  and in its finer environments (groups, binary and single systems) are well matched in the local universe down to the completeness limit of the observational sample, the model
over-predicts the number of low mass galaxies in the field at $z\sim0.6$ and in clusters at both redshifts. Above $M_\ast$=$10^{10.25}M_\odot$, it reproduces the observed similarity of the cluster and field mass functions, but not the observed evolution. Our results point out two shortcomings of the model:  an incorrect treatment of cluster-specific environmental effects and an over-efficient galaxy formation at early times (as already found by e.g. Weinmann et al. 2012). 
Next, we consider only simulations. Using also the \cite{guo11} model,  we find that the high mass end of the mass functions depends on halo mass: only very massive halos host massive galaxies, with the result that their mass function is flatter. Above $M_\ast$=$10^{9.4}M_\odot$, simulations  show
an evolution in the number of the most massive galaxies in all the environments. Mass functions obtained from the two prescriptions are different, however results are 
qualitatively similar, indicating that the adopted recipes to model the evolution of central and satellite galaxies still have to be better implemented in semi-analytic models.

\end{abstract}

\section{Introduction}
The $\Lambda$ Cold Dark Matter (CDM) hierarchical paradigm 
describes well the formation of large-scale structure in the Universe.
It is, however, difficult to explain all the observed trends for the galaxy populations in the context of this paradigm. This is, at least in part, due to the difficulties in treating relevant physical processes (e.g. feedback from supernovae and AGNs, stellar winds, etc) whose understanding is far from being complete.  

From an observational point of view, many studies have tried to quantify the role 
of the environment in shaping the physical properties of galaxies at different cosmic epochs.
In general,  denser environments host larger fractions 
of early-type galaxies \citep{hubble31}, 
that are typically more massive, redder, more concentrated, less gas-rich, and show lower star formation rates than late-type galaxies. 
(e.g., \citealt{kauffmann04, baldry06, weinmann06}).
These trends may be driven by 
the environment, 
and be the result of physical processes coming into play only after galaxies have become part of a structure like a group or a cluster,
or they might be mainly driven by intrinsic properties closely related to galaxy-intrinsic conditions (i.e. stellar mass) and be established 
beforehand, due to the fact that galaxy formation occurs at an accelerated rate in over-dense regions. 
As discussed in \cite{delucia12}, separating the two scenarios and differentiating their role in driving galaxy evolution  is  hard, since 
 they are strongly and physically connected.

In the last years, 
much attention has been focused on quantifying the environmental trends at fixed stellar mass (but see caveats pointed out in \citealt{delucia12}). In this paper, we will focus on one specific property of galaxy populations that is the galaxy stellar mass function. Since it results from a combination of hierarchical mass assembly of dark matter halos and different physical processes driving galaxy evolution (e.g., \citealt{dekel86, benson02a, wang08}), by comparing observational data at different cosmic epochs with theoretical predictions can provide important constraints on the entire galaxy formation processes. 

First studies focused mainly on field galaxies
(e.g., \citealt{fontana04, drory05, gwyn05, fontana06, borch06, bundy06, pozzetti07, 
pozzetti10, ilbert10, baldry12,ilbert13,muzzin13}). 
They showed that the number density of galaxies with $M_{\ast} \geq 10^{11}M_{\odot}$ exhibits 
relatively modest evolution from $z=1$ to $z=0$. 
This implies 
that the assembly of relatively-massive objects  is essentially complete
by $z\sim1$. On the other hand, the mass function of less massive 
galaxies evolves 
more strongly than that of massive ones, displaying a rapid increase of their number density from $z\sim1$ (but see \citealt{drory09}).
As pointed out by \cite{marchesini09}, however, a detailed analysis of random and (in particular) systematic uncertainties weakens significantly any claim on the mass-dependent evolution, particularly for the massive end. 

Recently, a few studies have started to investigate the mass function in galaxy clusters.
\cite{morph} 
found a quite strong evolution with redshift, consistent with the field evolution.  \cite{rosa2} (hereafter C13) and  \cite{mf_global} (hereafter V13) 
found that, at least above \M$\sim$10.25  at
$z=0$, and above \M$\sim$10.5 at $z=0.6$, the shape of the mass distribution  is very similar in clusters, groups and the field. 
Similar conclusions have been also extended to $z\sim1$ \citep{vanderburg13}.
 
V13 also showed that the evolution of the mass function in the interval $z=0.6-0.06$ is similar in clusters and in the general field.

On the theoretical side, semi-analytic galaxy formation models (e.g. \citealt{white91, kauffmann93, cole94, kauffmann99, somerville99, cole00, springel01, hatton03, baugh05,  delucia07, font08}), 
provide a powerful tool to interpret observational results in a cosmological context. 
However, it is important to keep in mind that usually models are normalized to fit a subset of low-$z$ observations and 
the field mass function has been often used by the most recent models as the primary constrain to tune the various model parameters. 

Several studies have  compared the observed field mass function to the one predicted by simulations. 
Semi-analytic models that include strong stellar feedback  reproduce well the $z = 0 $ mass function (e.g. \citealt{guo11} - hereafter G11, \citealt{bower12}), but
they struggle in reproducing  the mass function of  low-mass galaxies at higher redshift (\citealt{fontana06,marchesini09, drory09, fontanot09} - hereafter F09, \citealt{lofaro09, cirasuolo10}, G11). In particular, they overproduce low-mass galaxies at  $z> 0.5$, 
predicting almost no evolution in the number density of galaxies of mass $\sim10^{10} M_\odot$ since $z \sim 2$, 
in contrast with observational measurements for galaxies of the same mass whose number density is found to evolve by a factor $\sim$6 over the same redshift range (F09). In the models, low mass galaxies  are predicted to form too early and have too little ongoing star formation at later times (e.g. F09, \citealt{firmani10}, G11, \citealt{weinmann12}), so  their present day stellar populations are too old. 
  
For low-mass galaxies, it is generally believed that the discrepancies with observational data are due to an incorrect treatment of the star formation and stellar feedback process. In addition, as pointed out by \cite{weinmann12}, the problem is not limited to semi-analytic models but is also present in hydrodynamical simulations of galaxy formation. 
 
 A number of problems still affect model predictions also for the most massive galaxies (\M$>11$). Their evolution since $z\sim1$, which is driven by mergers, is marginally inconsistent with the observational results, being slightly too fast (F09).
 Models  also  underestimate both the number density and the star formation rate (SFR) of massive galaxies at $z>2$ (\citealt{marchesini09}, F09).  In addition,  models do not reproduce the observed chemical abundances: at the massive end, the predicted mass-metallicity relation turns over and is offset low with respect to the data (see, e.g., \citealt{delucia_borg12}).

The inaccuracy of the CDM paradigm seems to be related to the fact that model galaxies closely follow the
evolution of DM halos, 
while it is necessary to find a way to decouple the halo accretion rate and the star formation rate of 
galaxies \citep{weinmann12}.

Recently, some studies have  investigated the effect of assuming a Warm Dark Matter (WDM) power spectrum. \cite{menci12} argued that the WDM scenario may solve   the excess of   low-mass galaxies,  since it produces a smaller number of collapsed low-mass halos. 
However, \cite{kang13} asserted that 
the claimed success   might simply reflect a non-optimal parameterization of the physics of galaxy formation implemented in the model.  
This shows that a single observable (e.g. the stellar mass function) can not constrain the effects of the warm component on galaxy formation, even though accurate measurements of the mass function and the link between galaxies and DM halos down to the very low mass end can give very tight constraints on the nature of DM candidates.

All studies mentioned above focused on comparing model predictions with observational data for field galaxies. Very few studies have considered trends as a function of the environment (see e.g. \citealt{liu10} for an analysis of the local conditional stellar mass function). On the theoretical side, it is a well established results that the subhalo mass function depends only weakly on the mass of the parent halo (e.g. \citealt{delucia04,lee04, giocoli08}). It has to be considered, however, that galaxies and subhalos are not simply related \citep{gao04, sawala13}  so that a similar subhalo mass function does not necessarily imply a similar galaxy mass function. 

\bigskip
The aim of this paper is to  analyze what semi-analytic models predict for the mass function
at two different epochs ($z=0.06$ and $z=0.62$), in the field and in halos of different mass. 
Our aim is to test whether simulations are able to reproduce the observational 
results  
and eventually help improving the available models by analyzing in detail where they fail. 

The plan of the paper is as follows. 
Section 2 and 3 introduce the observational and theoretical samples, respectively. Section 4 describes the method used for our analysis, while Section 5 shows the basic comparison between the observed and predicted mass functions. In Section 6, we discuss results from simulations, and in Section 7 we summarize and discuss our results. Finally, Section 8 gives our conclusions. 

We assume $H_{0}=73 \, \rm km \, s^{-1} \,
Mpc^{-1}$, $\Omega_{0}=0.25$, and $\Omega_{\Lambda} =0.75$.  The adopted
initial mass function (IMF) is that of \cite{kr01} in the mass range 0.1--100
$M_{\odot}$. Magnitudes  are in the Vega system.

\section{Observations}
In this paper, we exploit four different observed samples to measure the galaxy stellar mass function of  low- and intermediate-$z$ field and cluster galaxies.

\subsection{Low-$z$}
We use the Padova-Millennium Galaxy and Group Catalogue (PM2GC - \citealt{rosa}), 
which is  a galaxy
catalog representative of the general field population in the
local Universe. It is a database built on the basis of the Millennium Galaxy Catalogue (MGC, \citealt{liske03}), a deep and wide B- imaging survey along an equatorial strip of $\sim  38\, {\rm deg}^2$. The  catalog contains only galaxies brighter than $M_B \leq-18.7$. 
By applying a friends-of-friends (FoF) algorithm, a catalogue of galaxy groups with at least three
members in the redshift range $0.04\leq z \leq$ 0.1 has been created (see \citealt{rosa}).  
Galaxies that after several iterations of the algorithm are within
1.5 $R_{200}$ from the group centre and 3$\sigma$ (velocity dispersion) from
the group redshift are considered  group members.
$R_{200}$ 
is defined as the radius delimiting a 
sphere with interior mean density 200 times the critical 
density of the universe at that redshift,
and is commonly used as an approximation of the group/cluster 
virial radius. The $R_{200}$  values for our structures are computed 
from the velocity dispersions using the formula  \citep{finn05}:
 \begin{small}
 \begin{eqnarray}\label{rm}
 R_{200}&=&1.73\frac{ \sigma}{1000 (km\, s^{-1})}\frac{1}{\sqrt{\Omega_{\Lambda}+\Omega_{0}(1+z)^{3}}}h^{-1}   (Mpc) 
 \end{eqnarray} \end{small}

Galaxies that do not satisfy the group linking
criteria adopted have been placed either in the catalogue of single
field galaxies, that comprises the isolated galaxies, or in the catalogue of binary field galaxies, that comprises the systems with two galaxies
within 1500 ${\rm km}\, {\rm s}^{-1}$ and 0.5 $h^{-1}$ Mpc. 

Stellar masses have been estimated following \cite{bj01} \citep{rosa}. 
Briefly, they were derived using the relation 
between  $M/L_{B}$  and  the dust-uncorrected rest-frame $(B-V)$ color
\begin{equation}\label{bj}
\log_{10}(M/L_{B})=-0.51+1.45(B-V)
\end{equation}
valid for a Bruzual \& Charlot model with solar metallicity and a \cite{salpeter55} IMF 
(0.1-125 $M_{\odot}$). Then,  they were converted  to a \cite{kr01} IMF, adding -0.19 dex to the logarithmic value of the masses.  The typical scatter of the mass uncertainties is $\sim 0.2-0.3 \, dex$  (see, e.g., \citealt{kannappan07}).

In this work, we refer to the galaxy sample described in C13: we consider galaxies at $0.04\leq z\leq 0.1$ in the field, in  groups, binary and single systems.  The mass completeness  limit  is $\log M_\star/M_\odot = 10.25$ and includes 1045 field galaxies.
 
The PM2GC covers  a much smaller sky area than the SDSS \citep{york00}, but is characterized by a better  imaging quality and higher spectroscopic completeness. In Fig.1 of their paper, C13 present a comparison between the PM2GC mass function and  literature results \citep{bell03, cole01, baldry12, li09}, and show a good agreement among the different estimates.  

\bigskip
For   clusters, we take advantage of the Wide-field Nearby Galaxy-clusters Survey (WINGS\footnote{http://web.oapd.inaf.it/wings} -  \citealt{fasano06}),
 a multi-wavelength survey  at \mbox{$0.04 < z < 0.07$}.
WINGS is based on deep optical (B,V) wide field images ($\sim35^\prime \times 35^\prime$) of 76 clusters. The clusters span a wide range in velocity dispersion (550$<$$\sigma$$<$1400 ${\rm km}\, {\rm s}^{-1} $) and X-ray luminosity (0.2$\times10^{44}$$<$${\rm L}_X$$<$5$\times10^{44} \,  {\rm erg} \, {\rm s}^{-1}$).
Besides the optical imaging data, a number of follow-ups were carried out to obtain additional homogeneous information for galaxies, such as optical spectroscopy \citep{cava09}, near- infrared (J, K) data 
\citep{valentinuzzi09} and U-band imaging \citep{omizzolo13}. An Omegacam/VST u, B and V follow-up of about 50  clusters is underway.

We use the  galaxy sample described in \cite{morph}. Briefly, we consider 21 clusters with 
spectroscopic completeness larger than 50\%.
The brightest cluster galaxies (BCGs) have been excluded, while only spectroscopically confirmed members within $0.6 R_{200}$ (the largest radius generally covered in WINGS clusters) have been included in the sample. 
Galaxies have been weighted for spectroscopic incompleteness using the ratio of the number of galaxies with a spectroscopic redshift to the number of galaxies in the parent photometric catalogue, as a function of galaxy magnitude \citep{cava09}.\footnote{The completeness is larger than 50\% for galaxies with $V<18$. This magnitude limit corresponds to a stellar mass of \M$\sim10.4$, below which the completeness correction becomes important.} Stellar masses have been estimated following \cite{bj01}  \citep{morph}. 
The mass complete sample includes all galaxies more massive than $\log M_\star/M_\odot =9.8$, for a total of 1229 galaxies.

We use the WINGS mass function  because, thus far, it is the only attempt to describe the galaxy stellar mass distribution  in clusters at $z\sim 0$. We note that  \cite{mercurio12} investigated the Shapley supercluster, but without explicitly isolating
clusters. Furthermore, the cluster mass function presented in \cite{baldry08} is based on masses determined from luminosities using a fixed M/L ratio for all galaxies.\footnote{The lack of studies of cluster stellar mass functions might be quite surprising.
An explanation for this might be that analyses focused mainly on conditional stellar mass function instead (see, e.g., \citealt{yang08}), which describes  the average number of galaxies as a function of galaxy stellar mass in a dark matter halo of a given mass. This approach allows to  disentangle the role of the galaxy formation model (which affects the relation between the mass of the halo and the galaxy stellar mass) from that of the adopted cosmology (which affects the number density), but strongly depends on the theoretical assumptions. }

\subsection{Intermediate-$z$}
For the field at $0.6 < z < 0.8$, we use the mass function presented in \cite{drory09} that is, among those published, the one with the lowest mass completeness limit. 
 The authors exploited the COSMOS  catalog with photometric redshifts
derived from 30 broad and medium bands described in \cite{capak07} and \cite{ilbert09}. They used galaxies with $i^+_{AB} < 25$, where the
detection completeness 
is $>90\%$ \citep{capak07}. 
They detected 36,885 galaxies at $0.6<z<0.8$ in an area of 1.73 $deg^2$.

\cite{drory09} derived stellar masses  comparing multi-band photometry to a grid of stellar population models of varying star formation histories (SFH), ages, and dust content (for further details, refer to \citealt{drory09}). They adopted a \cite{chabrier03} IMF (0.1-100 $M_\odot$). To reduce the uncertainties due to the different method adopted by \cite{drory09}, we apply a mean correction to the \cite{drory09} mass function, as described in Appendix \ref{drory_mass}.
Converting their stellar masses into our  IMF and cosmology, their mass completeness limit is $\log M_\star/M_\odot = 10$.

\bigskip
We also use the ESO Distant Cluster Survey (EDisCS - \citealt{white05}),
 a multi-wavelength photometric and spectroscopic survey 
of galaxies in 20 fields containing galaxy clusters ($400<\sigma<1100 \, {\rm km \, s^{-1}}$) and  groups 
($150<\sigma<400 \, {\rm km \, s^{-1}}$)  at 
$0.4< z <1$. Structures were 
selected as surface brightness peaks in smoothed images taken with a very wide optical filter ($\sim4500-7500$ \AA), and have high-quality multiband optical and near-infrared photometry \citep{white05} and spectroscopy \citep{halliday04, milvang08, spec}. 
Photometric redshifts were computed 
using two independent codes, a modified version of the publicly available Hyperz code \citep{bolzonella00} and the code presented in \cite{rudnick01, rudnick03, rudnick09}. 
Photo-$z$ membership  was established using a modified version of the technique first developed in \cite{brunner00} \citep{delucia04, delucia07, pello09}. 

Stellar mass estimates are presented in \cite{sfr, morph}; they have been determined following \cite{bj01}.
The photometric magnitude limit (I = 24) corresponds to a mass limit of $\log M_\star/M_\odot = 10.2$.

We use the EDisCS sample described in \cite{morph}, which includes all the photo-$z$ members of 14 clusters in the redshift range $0.4< z <0.8$, above the mass limit and within $0.6 R_{200}$.  BCGs have been excluded. The final sample consists of 489 galaxies. 
As discussed in the appendix of \cite{morph} and in V13,
  contamination from  photo-$z$ interlopers might alter the mass functions. However \cite{rudnick09} found 
that at the magnitudes used in this paper  the photo-z counts are fully consistent with the
statistically background subtracted counts.
Anyway, 
it should be noted that the stellar mass function measured using photo-$z$ membership becomes increasingly uncertain with decreasing magnitude/mass, where our data lack a good spectroscopic coverage. 
We exploit the spectroscopic sample to characterize the maximum mass reached by galaxies in clusters, since photo-$z$ interlopers could bias this estimate.

As for the local universe, no other cluster galaxy mass functions are available at these redshifts. 

\section{Theoretical predcitions} \label{theor_pred}
We take advantage of publicly available galaxy catalogues from semi-analytic models run on the Millennium Simulation \citep{springel05}. This uses  $10^{10}$ particles of mass
$8.6 \times  10^8 \, h^{-1} \, M_\odot$ to trace the evolution of the matter distribution in a cubic region of the Universe of $500\, h^{-1} \, Mpc$ on a side from $z = 127$ until $z = 0$, and has a spatial resolution of $5\, h^{-1} \, kpc$.  

We use two different semi-analytic models, to investigate
how different assumptions about 
the physical processes acting on the baryonic component
impact the evolution of the galaxy stellar mass function.

The semi-analytic model discussed in \cite{deluciablaiz07} (DLB07) builds on the methodology and prescriptions introduced in \cite{springel01, delucia04, croton06} and has been the first variant of the ``Munich'' models family that has been made publicly available.  The latest update is provided by the G11 model, that we use in the second part of this paper. 
Both models include prescriptions for supernova-driven winds, follow the growth of supermassive black holes and include a phenomenological description of AGN feedback. The G11 model is based on the DLB07 one, but it significantly differs in some areas.  
Several of the extensions involve a different treatment of some processes,  such as the strangulation of satellite galaxies (that is instantaneous after accretion in DLB07), and the supernova feedback
 (that is more efficient in G11);  and the introduction of some processes which were not previously included, such as the disruption of galaxies to produce intracluster light. In particular, when a central galaxy first becomes satellite, that is its halo is first linked to a more massive halo, G11 continue to treat it as a central galaxy, that is in the same manner as a galaxy at the centre of a main subhalo, until it falls within the virial radius of the centre of its new halo. At this point, they switch on tidal and ram-pressure stripping processes which can remove gas from the galaxy or even disrupt it completely.  This change leads to a reduction in the number of  satellite galaxies with respect to predictions from DLB07.  We refer to the original papers for more details.

Note that the DLB07 model was mainly constrained by the observed local K-band luminosity function.
The G11 model, on the other hand, was tuned to reproduce the observed stellar mass function in the local universe. 

\bigskip 
In the following, we refer to an output value of the simulations either as a ``sim-projected''   or a ``simulated'' quantity.
{\it Sim-projected quantities} (number of galaxies, velocity dispersion, etc.) are computed from the simulation with the same methods that would be used observationally and are projected on the $xy$ plane, while
{\it simulated quantities} are the 3D estimates provided by the simulation. 

As also discussed in \cite{poggianti10}, the former quantities can be compared to observational measurements. 

\subsection{The sim-projected sample}\label{sim_obs_sample}
To compare simulations to observations, we use only the DLB07 model and we assemble samples as similar as possible to the observed ones. The G11 model does not provide the magnitudes needed to compute the stellar masses as done with the data, and will be used only in the second part of this study.

To reproduce the observed clusters, using the available catalogues, we selected 150 halos  with $ 10^{13} \, M_\odot \leq M^{MS}_{200} \leq 10^{15} \, M_\odot$ at $z=0.06$ and 142 halos at $z=0.62$, uniformly distributed in mass.  $M^{MS}_{200}$ masses are computed from the N-body simulation as the mass enclosed within $R^{MS}_{200}$, the radius of a sphere which is centered on the most bound particle of the group and has an over density of 200 with respect to the critical density of the universe (the virial radius) at the redshift of interest.   
We then selected all galaxies within a box of 10 physical Mpc on a side, centered on each halo considered. 
3D velocity dispersions have been computed using all galaxies within $R^{MS}_{200}$ and more massive than $M_\star =  10^9 \, M_\odot$. 

To reproduce the observed field, we selected
portions  of ``simulated sky'' corresponding to square boxes of $\sim 38 \, deg^{2}$ ($30 \times 30$ Mpc) at $z= 0.06$, 
and to $\sim 1 \, deg^{2}$ ($17\times 17$ Mpc) at $z= 0.62$. At low-{\it z}, to match the observed field sample, the boxes extracted are 323 physical Mpc deep, while at higher redshift the maximum depth of the simulated box  ($500\, h^{-1} \, Mpc$) is used, which corresponds to  422 physical Mpc.
To account for cosmic variance, ten simulated field samples were selected at each redshift. No preselection on halo mass was applied to these boxes.

The model provides stellar masses and rest-frame Vega magnitudes that include the effect of the dust, in the \cite{buser78} system, 
 calculated using the models of \cite{bc03}. 
Instead of using the stellar masses provided by the models, we recomputed them using the \cite{bj01} formulation (Eq.\ref{bj}), as done for the observations. To this aim, we first converted the model magnitudes to the Johnson-Cousins system as defined in \citealt{bessell90}.\footnote{This is the photometric system needed to compute  masses using \cite{bj01}.}
A comparison between these stellar masses and those provided by the model is provided in Appendix \ref{model_mass}.

In the following, we consider a conservative mass limit of \M=9.4. 

The sim-projected quantities are computed using the $\sigma$ along the $z$-axis. Considering a different projection of the velocity dispersion or the average of the three does not considerably affect the results. 
Then, assuming the virial theorem is valid,\footnote{\cite{diemer13} 
argue that the Jeans equilibrium is a more accurate assumption for dark matter halos, that do not have a well defined boundary as assumed in the virial theorem.  However, here we adopt the same  equation used for observations.} the sim-projected $R_{200}$ 
can be estimated from the velocity dispersion as given in Eq.\ref{rm}. We note that \cite{poggianti10} showed that both sim-projected velocity dispersions and sim-projected  $R_{200}$ similarly deviate from the velocity dispersion obtained from the halo mass and from the theoretical $R^{MS}_{200}$ radius, underestimating them at low values and overestimating them at high values (see also \citealt{biviano06}).

Our sim-projected samples are defined as follows:
\begin{description}
\item{{\it Field}: 
we consider galaxies from the snapshot corresponding to $z=0.06$ and $z=0.62$ respectively. 
To take into account the cosmic variance (or, more accurately,  sample variance), we compute a mean 
of the ten representations of the fields and consider the scatter among them in the errors.   Above the PM2GC (COSMOS) mass limit, we are left with 
a mean value of 1439 (5093) galaxies. } 
\item{{\it Clusters}:
at $z=0.06$ we consider the 47 halos with
$550<\sigma<1400 \, \, {\rm km \, s^{-1}}$, and the 77 halos at $z=0.62$ halos with
$400<\sigma<1100 \, \,{\rm km \, s^{-1}}$.\footnote{The velocity dispersion distribution of simulated clusters is consistent with that of the observed ones, so that trends can not be driven by the size of the clusters.} 
To measure the mass functions, from the sim-projected samples, we extract randomly 21 clusters at low-$z$ and 14 clusters at intermediate redshift. We repeat the sampling ten times, to take into account the cluster variance. 
Galaxies
whose velocity $v_z$
is within $3\sigma$ 
are considered members of the halos. We note that this is much larger than the size of the boxes we extracted: at low-$z$ $3\sigma$ spans from 20 to 50 Mpc, at higher $z$ from 10 to 25 Mpc. 
For the mass functions, we exclude the central galaxy of each halo and take into account only galaxies within the projected $0.6 R_{200}$, where $R_{200}$ has been computed from the $\sigma$ as in Eq. \ref{rm}. Above the WINGS (EDisCS) mass limit, we are left with a mean value of  
2877 (532) galaxies.} 
\end{description}

\subsection{The simulated sample}\label{sim_sample}
\begin{table}
\caption{Number of galaxies with \M$>$9.4 in the different halos and at different redshifts, for both models. \label{numbers}}
\centering
\begin{footnotesize}
\begin{tabular}{ll|cc|cc}
 & & \multicolumn{2}{|c|}{{\bf DLB07}}				 & \multicolumn{2}{c}{{\bf G11}} \\

	&& $z$=0.06 & $z$=0.62 & $z$=0.06 & $z$=0.62 \\
\hline
\multirow{3}{*}{$M_{halo}\sim 13.4$}	& $r/R_{200}<0.6$ &325448&316028&291311&305035\\
							&$r/R_{200}<1$& 478421&462758&405084&426654\\
							&$1<r/R_{200}<3$&256045 &191582&194560&162771\\
                                   
\hline
\multirow{3}{*}{$M_{halo}\sim 14.1$} & $r/R_{200}<0.6$ &109044&73776&100834&72542\\
							&$r/R_{200}<1$&171131&115355&149844&108260\\
							&$1<r/R_{200}<3$& 97718&47764&76842&41642\\
\hline
\multirow{3}{*}{$M_{halo}\sim 15.1$}	& $r/R_{200}<0.6$ &14608&2519&100834&2406\\		
							&$r/R_{200}<1$&23653 &3909&20662&3619\\
							&$1<r/R_{200}<3$&11419 &1555&9128&1405\\
\hline
$field$ 						&		&5460843&5199001&4196571&4332813\\
\end{tabular}
\end{footnotesize}
\end{table}

When analyzing only simulations,  
we take advantage of all the information coming from the Millennium Simulation and use  3D values. We exploit both the DLB07 and G11 models.

We extract
all galaxies from the snapshot corresponding to $z=0.06$ and $z=0.62$ respectively, to characterize the field. Both at low- and intermediate-{\it z}, we also  extract all halos in three different halo mass bins: 76853   and  62333 halos with $13.25<\log M_{halo} <13.55$ (least massive halos), 2981  and 1910 halos with $13.9<\log M_{halo} <14.25$ (intermediate massive halos) and  41 and 6 halos with $14.9<\log M_{halo} <15.25$ (most massive halos), where $\log M_{halo}=\log(M_{200}^{MS}/M_\odot)$, respectively. 
As done for the sim-projected sample, for each halo we extracted 
 a box of 10 physical Mpc on a side, centered on the most bound particle. 
3D velocity dispersions have been computed using all galaxies within $R^{MS}_{200}$ and more massive than $M_\star =  10^9 \, M_\odot$.  

To compute the galaxy stellar mass function, we have considered all galaxies that are members, and excluded central galaxies. 
Finally, we consider separately  galaxies at different  halo centric distances: within 0.6$R_{200}$, within 1$R_{200}$ and within 1-3$R_{200}$. 

The number of galaxies in the different samples are given in Tab.\ref{numbers}. The G11 model always has a lower number of galaxies than the DLB07 one, at any redshift and in any environment.   
This is just due to the stronger stellar feedback assumed in the G11 model, that reduces the number of low-mass galaxies.

For these samples, we use the stellar masses provided by the models,  converted to a  \cite{kr01} IMF.

\section{Methods}\label{method}
To quantify the similarities between the mass functions in different environments, 
we  analyze several aspects of the mass distributions.

(1) We 
 inspect the shape of the data distributions.
 We consider bins of 0.2 dex in stellar mass and express the mass functions in units of comoving volume 
(in units of number per $h^{-3} \,  Mpc^3 
dex^{-1}$).  To compute the volume for clusters, we consider  the sphere (or part of the sphere for observations and sim-projections), defined in terms of virial radius, which includes galaxies in the considered sample.
Details on how clusters mass functions have been normalized can be found in Appendix \ref{app_normalization}.
In all the figures, we do not plot errors along the $x$-direction, but errors on stellar masses are typically of 0.2-0.3 $dex$.
Errors along  the $y-$direction  are computed adding in quadrature the Poissonian errors \citep{gehrels86} and  the uncertainties due to cosmic variance or cluster-to-cluster variations, computed from the sim-projections. 
Following \cite{marchesini09}, we measured $\phi$ separately for each of the $n$ fields or clusters, and then we estimated the contribution to the error budget of $\phi$ from cosmic/cluster variance using
$\sigma_{cv} = rms(\Phi_i)/\sqrt{n}$. The values obtained are then used also for observations. 
When considering only simulations, neither cosmic nor cluster variances are considered, since we are using $all$ the available galaxies.

(2) 
We use the Kolmogorov-Smirnov (K-S) test, that quantifies the probability  that two data sets are drawn from the same parent distribution.\footnote{We note that the K-S test does not take into account the errors.} A ``low probability'' ($P_{K-S} < 5\%$)  
means that two samples are different; on the contrary a ``high probability'' means  that the test is unable to find differences. In the following figures, only significant K-S results will be given. For the WINGS sample, we consider completeness weights when running the test. 
When we investigate only simulations, we do not use the test, since, given the very large numbers of galaxies, it turns out to be meaningless.

(3) We
consider analytical fits to the mass functions using a Monte Carlo Markov chain method. 
Assuming that  the number density $\phi(M)$ of galaxies 
can be described by a \cite{schechter76} function, the mass function can
be written as
\begin{small}
\begin{equation}\label{eq:sc}
\phi (M) = (\ln10) \, \Phi^*  10^{(M-M^*)(1+\alpha)} \, \exp\left(-10^{(M-M^*)}\right)
\end{equation}
\end{small}
where $M = \log (M_\star/M_\odot)$, $\alpha$ is the low-mass-end slope,
$\Phi^*$ the normalization and $M^*  = \log (M_\star^* /M_\odot)$ is 
 the characteristic mass. 
Schechter function fits are computed   only  above the   completeness limits.

\begin{figure*}
\centering
\includegraphics[scale=0.44]{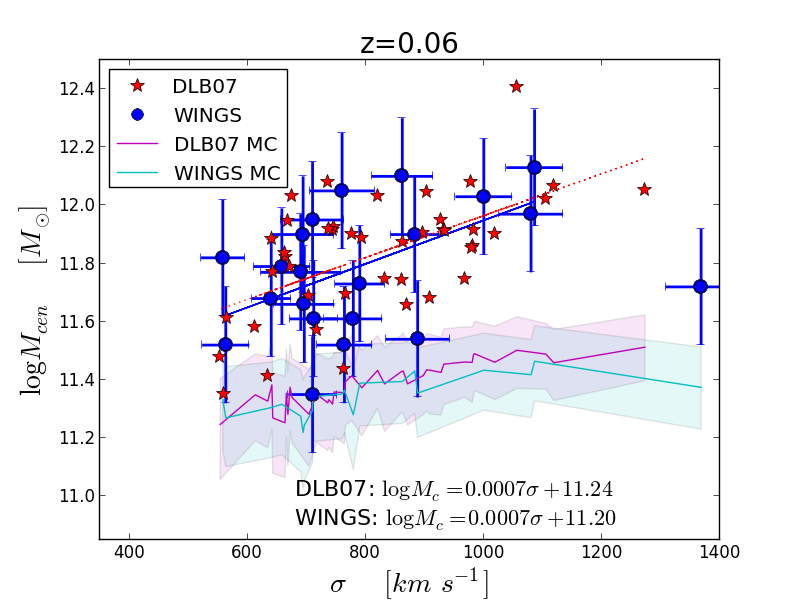}
\includegraphics[scale=0.44]{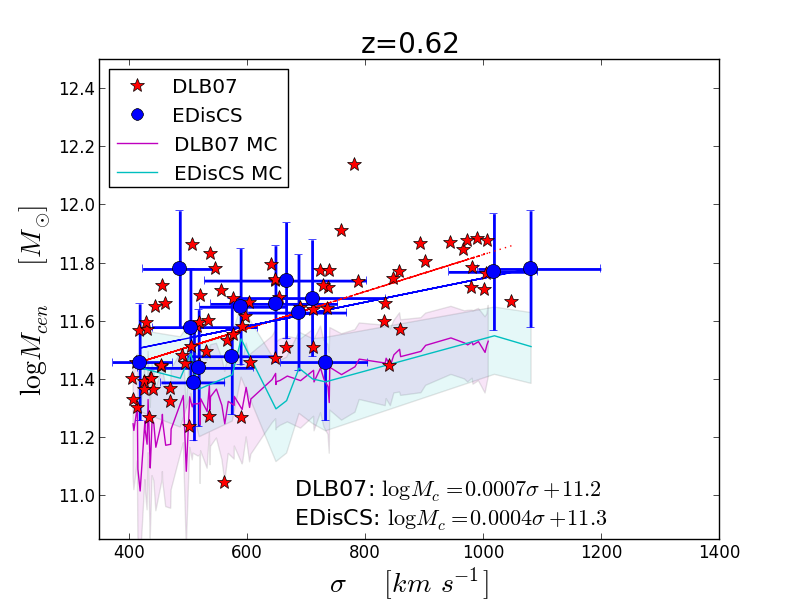}
\caption{$M_{cen}$-$\sigma$  for observed (blue points) and  sim-projected (red stars) structures, at $z$=0.06 (left panel) and $z$=0.62 (right panel).  Galaxies of all the sim-projected halos are plotted.  Solid blue and dotted red lines are the fit to the data. Magenta and cyan regions represent  the Monte-Carlo estimates for the most massive galaxy as a function of cluster velocity dispersion (see text for details). \label{Mhalo_Mcen}}
\end{figure*}

Our Schechter fits are used to characterize  mass functions, but the procedure
does not take into account uncertainties on the estimates of the stellar mass.
However, errors and uncertainties associated with masses are not negligible.
E.g. \cite{marchesini09} showed that 
both systematic and random errors can arise from the unknown true SFHs, metallicities and dust corrections, 
and also from photometric redshift errors, differences in stellar population models, the unknown stellar IMF and its evolution, and cosmic variance. 
The error convolution can have a significant impact upon the shape of the high-mass end of the mass distribution (e.g., F09 and references therein).

In addition, the high mass end is affected by the \cite{eddington13} bias: 
in the mass range where the number counts are steep, 
more objects of intrinsic low mass will be scattered to higher values of stellar mass than systems of intrinsically high mass being scattered to lower masses. This results in an over-estimate of the true galaxy counts at high stellar masses.

As a consequence, when we use stellar masses determined using \cite{bj01} (\S 5),
we should take into account the impact of errors on the  mass estimates. 
We assume that the error has a Gaussian distribution (independent of mass and redshift).
Following an approach similar to that described in \cite{behroozi10}, we 
perform additional fits to our mass functions, with the aim of estimating the Schechter parameters that describe 
the ``intrinsic'' mass function ($\phi_{true}$). The observed mass function, convolved with the errors, is:
\begin{small}
\begin{equation}\label{eq:sc_cor}
\phi(M) = (\ln10)\int{\phi_{true}(m^\prime)\frac{1}{\sqrt{2\pi}\sigma} \exp{\frac{-(\ln{10^M}-m^\prime)^2}{2\sigma^2}}dm^\prime}
\end{equation}
\end{small}
where $m = \ln (M_\star/M_\odot)$, and $\sigma$ is the uncertainty on the stellar mass, assumed $\sim 0.2 \, dex$.
The effect of the convolution is:
\begin{small}
\begin{eqnarray}\label{eq:sc_corfin}
\phi(M) &=& (\ln10) \frac{1}{\sqrt{2\pi}\sigma} \int{{\Phi^{\prime}}^*\exp{ [ (\alpha^\prime+1)\left ( m^\prime-{\mathcal{ M^{\prime}}^* }\right ) ]} \times} \nonumber \\
&\times& \exp{[-\exp{(m^\prime-{\mathcal{ M^{\prime}}^* })}]}  \exp{[\frac{-(m^\prime-\mathcal{ M^{\prime}}^*)^2}{2\sigma^2}]} dm^\prime
\end{eqnarray}
\end{small}
where $\Phi^{\prime}$, $\alpha^\prime$ and ${\mathcal{ M^{\prime}}^* } = \ln (M_\star/M_\odot)$ are the best-fit  parameters we can insert in  the Schechter function
to get  $\phi_{true}$.

In our figures, we  report  fits  of $\phi_{true}$ to give an indication of how much the mass functions change when comprehensive error estimates are taken into account. 
However, in the comparison between observations and sim-projections, since we treat simulations in the same way as observations, we simply use $\phi(M)$, because this is the approach commonly used in observational studies (e.g., \citealt{bundy06, pozzetti10}). Our choice is also justified by the fact that errors inherent to the transformation from magnitudes
to masses obtained with spectrophotometric models (which dominate the total error) are common to both observations and sim-projections in our approach.

(4) We also 
analyze the highest
mass reached by galaxies in each environment, hereafter ``maximum mass'', both including and excluding  BCGs  in clusters. This  is another important aspect that can help us to 
characterize the environment-mass relation.

\section{Results: Comparison between sim-projected and observed mass functions}

In this Section we use the samples described in \S2.1, \S2.2 and \S\ref{sim_obs_sample}. 

We begin with investigating central cluster galaxies, that are excluded from our mass function analysis. Both in simulations and observations, ``total'' magnitudes are used to compute stellar masses, hence they may include some Intra Cluster Light.
Figure \ref{Mhalo_Mcen} shows that the mass of the central galaxy correlates with the velocity dispersion\footnote{Using the halo mass given by simulations instead of the velocity dispersion gives similar results.} of the hosting halo, at both redshifts considered. Observations and sim-projections\footnote{For sim-projections we use the central galaxy of all halos with 
$550<\sigma<1400 \, \, {\rm km \, s^{-1}}$ at low-$z$ and with
$400<\sigma<1100 \, \,{\rm km \, s^{-1}}$ at high-$z$.} are  in agreement, showing similar slope of the relations. This is a sign that the environment has a strong effect on the mass of the central galaxy. 
This result resembles that presented for EDisCS' BCGs in \cite{whiley08}: clusters with large velocity dispersions tend to have  BCGs with larger stellar masses. 
The strong link between central galaxies and hosting halo is known, and has been discussed by several authors (e.g.,   \citealt{shankar06, wang06, moster10, leauthaud12}). 

\begin{figure*}
\centering
\includegraphics[scale=0.44]{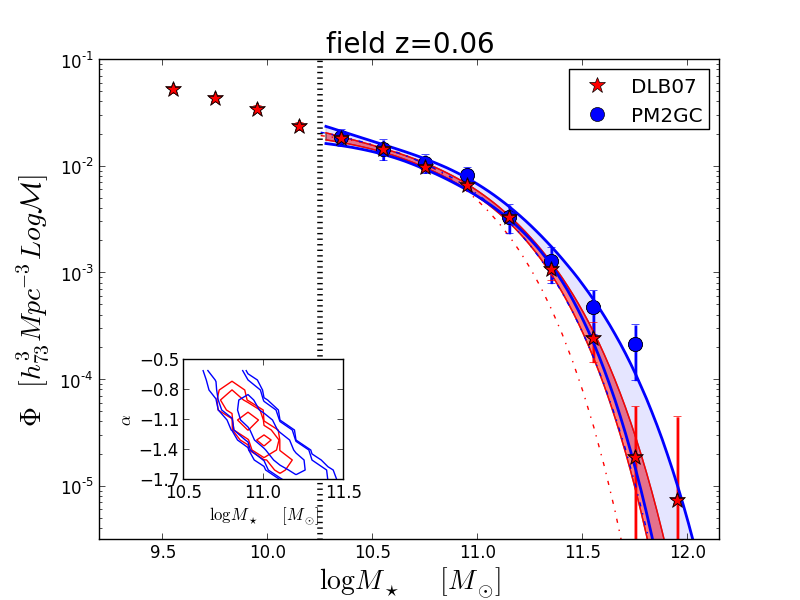}
\includegraphics[scale=0.44]{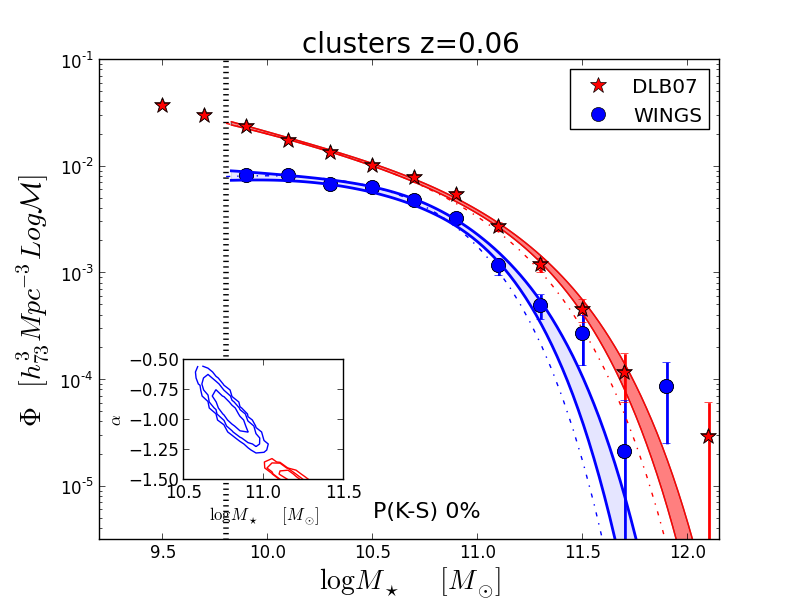}
\includegraphics[scale=0.44]{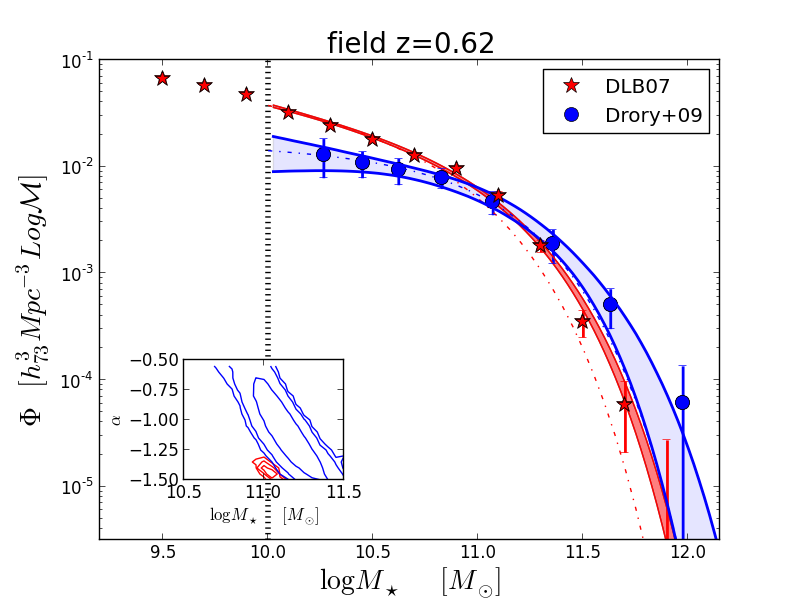}
\includegraphics[scale=0.44]{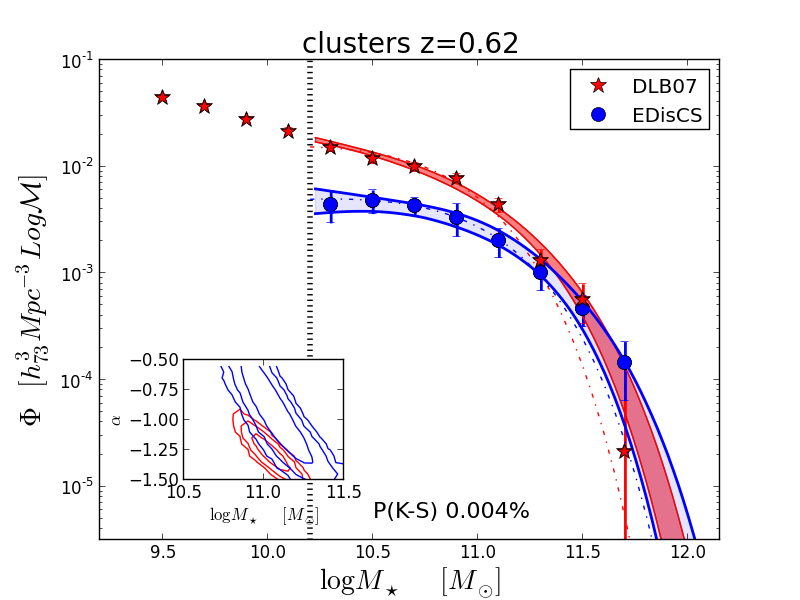}
\caption{Observed (blue symbols) and  sim-projected (red symbols) mass functions, for galaxies in the field and in clusters at low and intermediate-$z$ respectively. Mass functions are normalized to the comoving volume probed by the samples. 
Errorbars on the y-axis are computed combining the Poissonian errors \citep{gehrels86} and the uncertainties due to cosmic variance or cluster-to-cluster variations. For the \cite{drory09} sample, errors include also the uncertainty related to  photometric redshifts.
Points and crosses represent the measured mass functions, solid lines and shaded areas represent Schechter fits with $1\sigma$ errors, dashed lines represent the mass function ($\phi_{true}$), obtained from the deconvolution with the uncertainties on masses (see text for details).  The K-S probabilities are also shown as percentages. The bottom left insets in each panel show the $M^*$ and $\alpha$ Schechter fit parameters with $1, 2, 3 \sigma$ errors contours. The dotted vertical line shows the mass limit of each sample.\label{cfr_simobs_obs}}
\end{figure*}
The detected trend of more massive galaxies in higher velocity dispersion halos may be attributable to the environment. However, there is a statistical effect which can cause such a trend. If stellar masses are randomly drawn from the  mass function of galaxies, then massive halos, that host a larger number of galaxies, have a higher probability to host more massive galaxies \citep{tremaine77, bhavsar85, lin10, dobos11, paranjape12, more12}. To disentangle the statistical effect from the environmental effect, we adopt the following approach. For each cluster, we randomly sample $N$ galaxies from the observed field stellar mass function at similar redshift, where $N$ is the number of galaxies (including the BCG) in that cluster. 
We perform the random sampling 1000 times in order to sample the probability distribution of the maximum mass. The results are plotted in  Fig.\ref{Mhalo_Mcen}.
At both redshifts, the random sampling agrees quite well with the data for low-mass structures, while there is a deviation for the most massive halos. In this regime, at any fixed velocity dispersion, the observed and sim-projected masses are systematically higher than those coming from the random sampling. The effect is more noticeable al low- than at higher $z$. 
This test  suggests that the relation can indeed be ascribed to the environment, in particular  for very massive structures (in agreement with the results presented in \citealt{rodriguez13}).

\begin{table*}
\caption{Best-fit Schechter function parameters ($M_\star^{*}$, $\alpha$,  $\Phi^{*}$) for the mass functions of galaxies for different environments and redshifts. Fits are computed only above the completeness limit of each sample.\label{tab:fit_simobs}}
\centering
\begin{footnotesize}
\begin{tabular}{ll|ccc|ccc}
&	&\multicolumn{3}{|c|}{{\bf observations}}				 	& \multicolumn{3}{c}{{\bf DLB07 sim-projections}} \\
\hline
&	&$\log (M_\star^*/M_\odot) $ & 	$\alpha$ &$ \Phi^{*} $						&$\log (M_\star^*/M_\odot) $ & 	$\alpha$ &$ \Phi^{*} $\\
\multicolumn{2}{l|}{field  $z\sim0.06$}		& 11.1$\pm$0.2		&-1.4$\pm$0.3   &0.005$\pm$0.003		&10.97$\pm$0.06  &-1.3$\pm$0.1 &0.006$\pm$0.002  \\
&		groups  							& 10.9$\pm$0.1		&-1.$\pm$0.2   &0.0025$\pm$0.0007		&10.9$\pm$0.1  &-1$\pm$0.2 &0.0025$\pm$0.0007  \\
&		binary systems  					& 10.9$\pm$0.2		&-0.9$\pm$0.5   &0.0013$\pm$0.0006		&11$\pm$0.1  	&-1.5$\pm$0.4 &0.0007$\pm$0.0005  \\
&		single galaxies 					& 10.8$\pm$0.2		&-1.2$\pm$0.4   &0.003$\pm$0.001		&10.5$\pm$0.1	&-0.5$\pm$0.3   &0.005$\pm$0.001  \\
\multicolumn{2}{l|}{clusters $z\sim0.06$}	& 10.61$\pm$0.07	&-0.7$\pm$0.1   &0.0069$\pm$0.0009 		&11.17$\pm$0.05 &-1.44$\pm$0.04 &0.0027$\pm$0.0004  \\
\multicolumn{2}{l|}{field $z\sim0.6$}		& 11.1$\pm$0.06 	& -1.0$\pm$0.3 	& 0.006$\pm$0.002 		&10.99$\pm$0.03 &-1.43$\pm$0.04&0.0070$\pm$0.002  \\
\multicolumn{2}{l|}{clusters $z\sim0.6$}	& 11.0$\pm$0.1		&-0.8$\pm$0.3   &0.003$\pm$0.001		&11.01$\pm$0.08 &-1.3$\pm$0.1 &0.006$\pm$0.001  \\
\end{tabular}
\end{footnotesize}
\end{table*}

We now focus only on cluster satellites and  field galaxies and investigate whether simulations can reproduce the observed mass function in different environments and at different redshifts.  We stress that comparisons make sense only above the  mass limit of each observed sample.   
The fits presented  in the figures are computed  above the observed mass limit, hence comparisons between samples with different completeness limits are not meaningful.

As shown in Figure \ref{cfr_simobs_obs} and in Table \ref{tab:fit_simobs},\footnote{The values listed  are not the same as those given in  \cite{morph}, C13, V13,    for the same samples due to the different cosmology, normalization adopted and fitting routine. However, the reported values are in agreement within the errors with those already published. 
\cite{drory09} use a double Schechter to fit data, so their best-fit values are not directly comparable to ours.} the agreement between sim-projections and observations is only partial.  

At low-$z$, in the field  (upper left panel), sim-projected  and observed mass functions are very similar  ($M^*$ and $\alpha$ are  in agreement), as also pointed out by the K-S test. Sim-projections  reproduce   the observed number density, at any mass above the limit. On the other hand, they do not predict the right maximum mass reached by galaxies in the observed field: in the sim-projected sample there are galaxies with \M$\sim11.9$ that instead are missing in the observations. 

At similar redshifts in clusters (upper right panel), sim-projected and observed  mass functions are different: the  WINGS mass distribution lies always below the sim-projected one, indicating that it has less low-mass galaxies than the DLB07 model. For  \M$\leq10.5$,  the WINGS mass function becomes flat, while the sim-projected one keeps rising. At high masses, the sim-projected mass function 
extends to higher masses than the observed mass function.
Both the K-S test and the  Schechter fit confirm these visual findings:  in the sim-projected sample $M^*$ is much larger than in WINGS, while $\alpha$ is smaller (parameters not compatible at 3-$\sigma$ level).

\begin{figure*}
\centering
\includegraphics[scale=0.44]{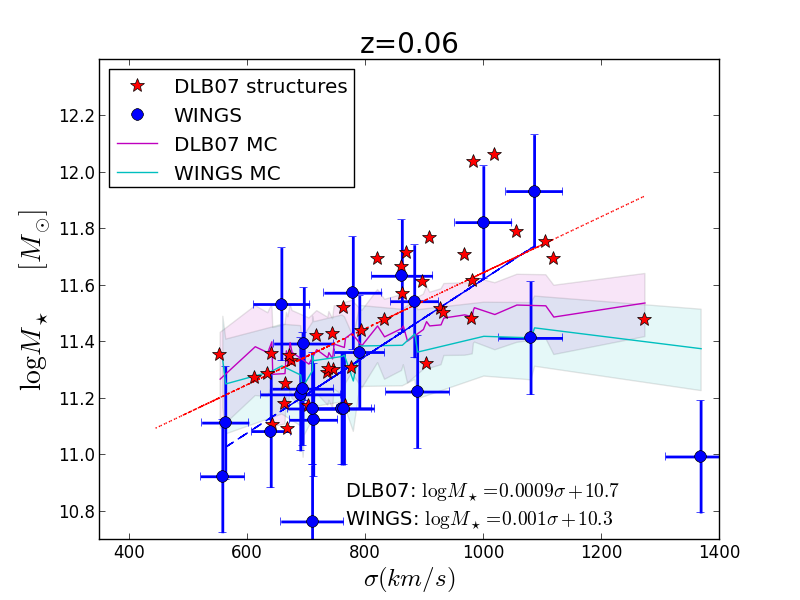}
\includegraphics[scale=0.44]{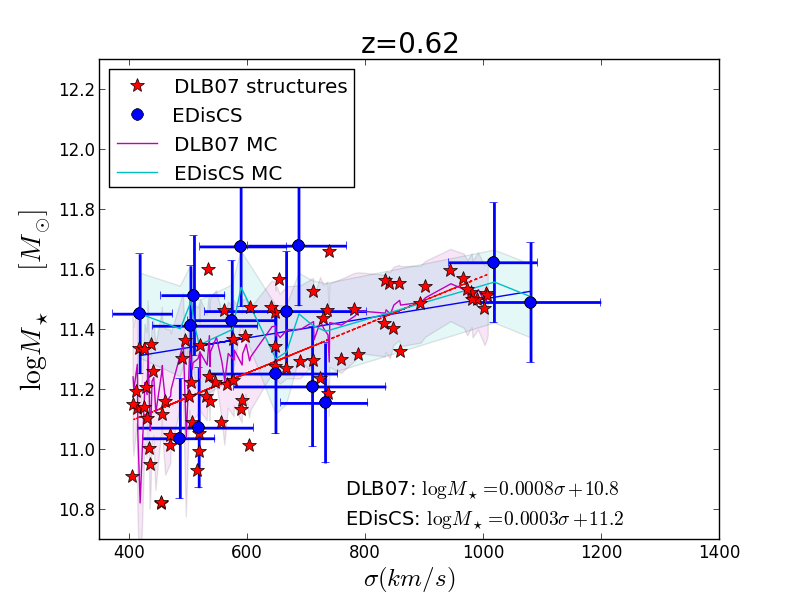}
\caption{$M_{max}-\sigma$ relation for observed (blue symbols) and  sim-projected (red symbols) structures, at z=0.06 (left panel) and z=0.62 (right panel). $M_{max}$ is the most massive galaxy in each structure (excluding the BCGs). For EDisCS, the spectroscopic sample has been used,  to avoid the contamination from interlopers in the photo-$z$ sample. Magenta and cyan symbols represent  the Monte-Carlo estimate of the mass of the most massive galaxy in such clusters (see text for details). \label{Mmax_sigma}}
\end{figure*}

At higher $z$,  simulations do not reproduce well the observed field mass function (bottom left panel). At \M$<$11,  the sim-projected mass function rises more steeply than the observed one, while at higher masses 
there are hints of an excess of very massive galaxies in the observed mass function compared to the sim-projected. 
However, we remind that in this regime  
uncertainties are large and also the correction we applied to the \cite{drory09} mass function may not be very accurate at these masses.
The results of the Schechter fit confirm that the two samples are not drawn from comparable distributions.\footnote{We can not perform the K-S test since we do not have at our disposal the unbinned counts from \cite{drory09}.}
Our results are in agreement with those discussed e.g. in F09: comparing three different semi-analytic models to observations, they found that for \M$<11$ all the models overpredict the observed  mass function, with the discrepancy increasing with increasing redshift. 
At similar redshifts, sim-projected and observed clusters (bottom right panel) show different shapes of the mass functions. 
We note that EDisCS is the only non spectroscopic sample, 
and, although as discussed in \S2.2 the contamination from  photo-$z$ interlopers does not alter the shape of the mass function,  
 the number density is influenced by it. 
To account for this,  we correct the EDisCS number density using the ratio of the counts in the  photo-$z$ sample to the counts weighted for incompleteness  in the  spectroscopic sample, in the mass range across which they overlap. This factor is independent of mass and it is equal to 2.8.\footnote{This renormalization does not influence the shape of the mass functions.}  
At \M$<11$, 
the EDisCS mass function is flat, while the DLB07 one keeps rising. The K-S test 
gives a high probability that the two distributions are different. 
At high masses, the two mass functions are comparable and the reached 
 maximum mass is similar. The Schechter fits are different at 1-$\sigma$ level. 

In all the panels, we also show  $\phi_{true}$ (see \S\ref{method}). This shows how statistical errors on individual stellar masses result in a more extended high mass end than the real one. 

\begin{figure*}
\centering
\includegraphics[scale=0.32]{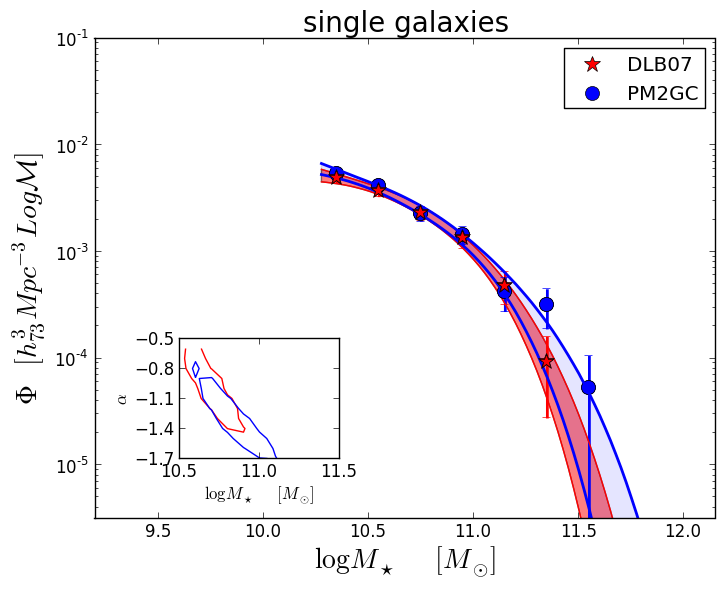}
\includegraphics[scale=0.32]{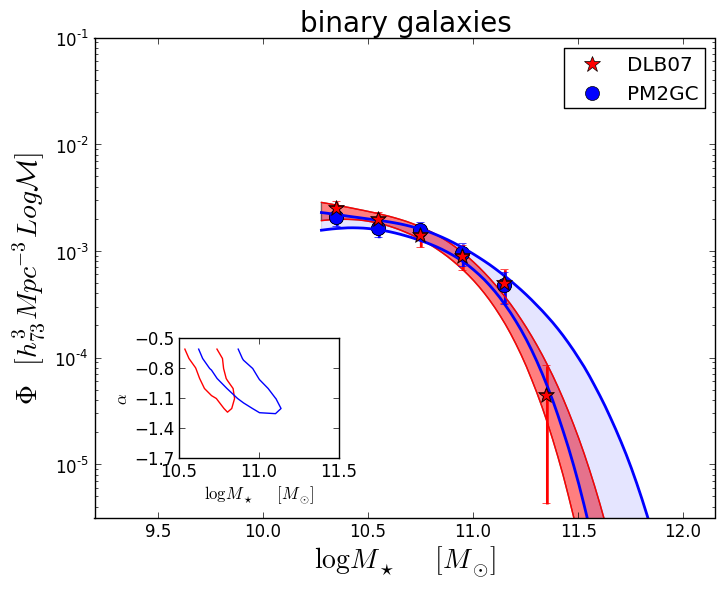}
\includegraphics[scale=0.32]{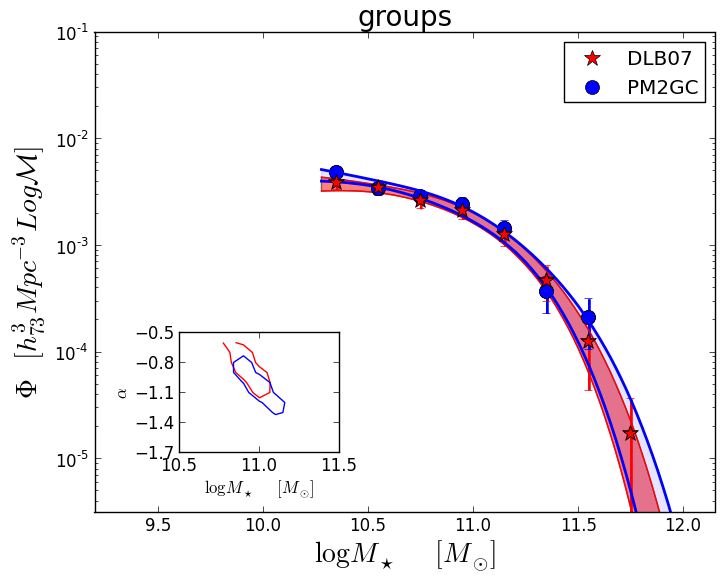}
\caption{Observed (blue symbols) and  sim-projected (red symbols) mass functions, for galaxies in the finer environments in the field at low-$z$. Normalization, errorbars, labels and inset are as in Fig.\ref{cfr_simobs_obs}. In the insets, only the 1-$\sigma$ contour are shown.\label{cfr_simobs_obs_field}}
\end{figure*}

Figure \ref{Mmax_sigma} shows the mass of the second most massive galaxy 
as a function of the cluster velocity dispersion. 
While BCGs in massive clusters considerably deviate from the trends found for the entire galaxy population, 
the second brightest galaxies are simply the statistical extreme of such population \citep{lin10, shen13} and it is thus interesting to consider
whether their mass correlates with halo mass.
For EDisCS, we use the spectroscopic sample, to have a more reliable indication on the cluster members.  
The agreement between observations and sim-projections is very good. At both low-$z$ (left panel) and intermediate-$z$ (right panel) and for both sim-projected and observed cluster samples  there is a strong  dependence of the second most massive galaxy on the mass of the halo: the more massive the halo, the larger is the stellar mass of the second most massive galaxy. 
Comparing the trends at the two redshifts, we find that the relation is shifted toward slightly higher masses at low redshift, suggesting that in the local universe structures host more massive galaxies.

As for the central galaxies, the detected trend 
might be attributable to the environment. 
We  perform the same random sampling describe above to  disentangle the statistical effect from the environmental effect. In this case we exclude the central galaxies from the samples. The results are plotted in  Fig.\ref{Mmax_sigma}.
At low-$z$, both for the observed and sim-projected clusters, the random sampling predicts a flatter relation 
than that obtained using the real samples, suggesting that the environment is  playing a role in the most massive halos. In contrast, at higher redshift, smaller differences between the randomly sampled galaxies and real samples are detected, so that at this redshift the relation found can be explained by a purely statistical effect.  A larger number of halos with high velocity dispersion would be useful to place stronger constraints on this finding.

\subsection{A more careful analysis of the field mass function}\label{field}

In the previous section, 
we have shown how well simulations reproduce the general field  (meant as a wide portion of the sky, including all environments) mass function in the local universe. Since the general field sample is the sum of group, binary systems and single galaxies, we aim to understand if simulations are able to reproduce the galaxy mass function of this finer division of environments.
To do this, we applied to the sim-projected field samples the same FoF we used for the PM2GC (see \S2 and \citealt{rosa}) and obtained samples that describe the single galaxies, binary systems and galaxies in groups. From both sim-projections and observations, we excluded galaxies in groups  with a
velocity dispersion $\sigma > 500 km/s$, to eliminate  
a possible contamination from clusters.

Figure \ref{cfr_simobs_obs_field} shows
the comparison between the sim-projected and observed mass function in the three different environments. The agreement is striking: in any given environment mass distributions are indistinguishable. The result is also supported by the analysis of the Schechter fits (see Tab.\ref{tab:fit_simobs}). 

\begin{figure*}
\centering
\includegraphics[scale=0.45]{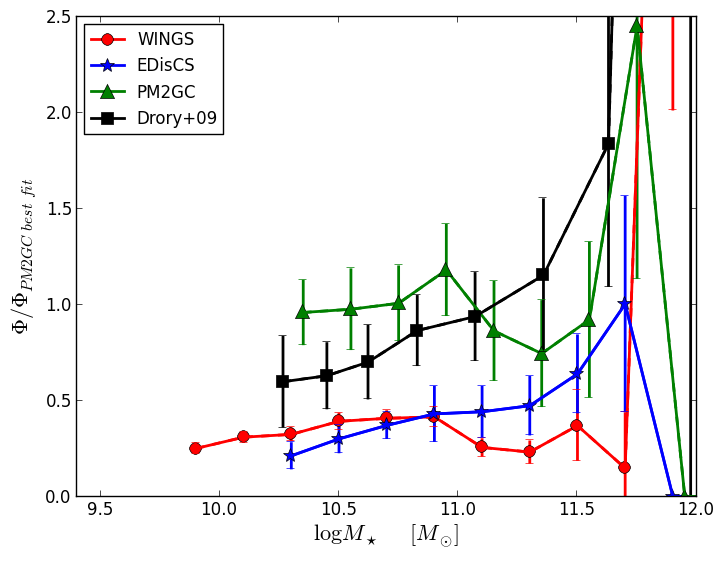}
\includegraphics[scale=0.45]{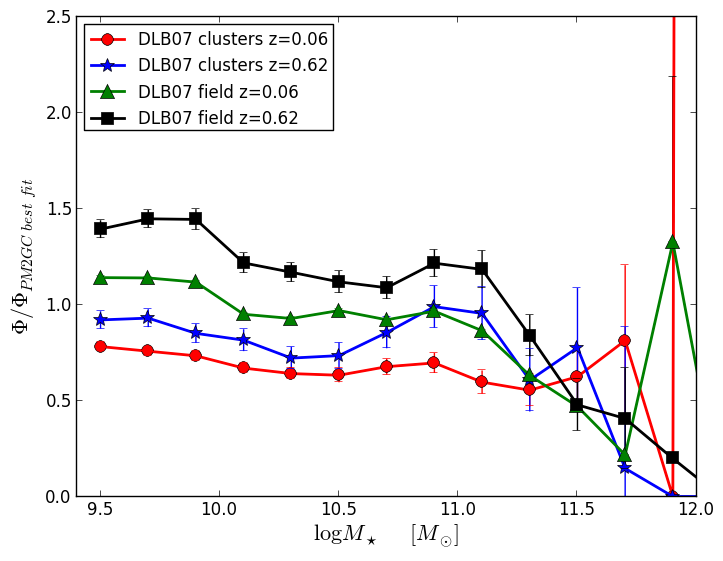}
\caption{ Ratio of the observed (left panel) and  sim-projected (right panel) mass functions to the PM2GC best fit Schehcter function, 
 for galaxies in clusters (red: low-$z$, blue: intermediate-$z$) and in the field (green: low-$z$, black: intermediate-$z$). Errorbars on the $y-$axis are obtained by error propagations.
\label{cfr_simobs_obs_evol}}
\end{figure*}

Hence, the semi-analytic model correctly includes all the processes that are responsible for the assembly of galaxies in these finer environments, groups included. This finding is important above all in the light of the fact that the cluster mass function at the same redshift is not reproduced, suggesting an inaccuracy in the treatment of cluster specific processes, as it will be discussed later on, in \S\ref{cl_gr}.

\subsection{Dependence on the environment and evolution}\label{evol}
The two main results of C13 and  V13 were that 
(1) both at low- and intermediate-$z$, the {\it shape} of the mass distribution of galaxies in clusters and in the field is similar, (2) the  
evolution of the shape of the mass function from $z \sim 0.6$ to $z \sim 0.06$  is the same in clusters and field. 
We now wish to test whether samples drawn from simulations are  able to reproduce these observational findings.

In Figure \ref{cfr_simobs_obs_evol}  we plot the ratio of the observed and sim-projected mass functions to the best fit Schechter function of our observed field mass function in the local universe (PM2GC) ($\Phi/\Phi_{PM2GC - bf}$), used for reference.\footnote{The choice of adopting the Schechter fit of the PM2GC data and not the data itself entails that the PM2GC line does not trace y=1 for all masses. }
This is another way of detecting differences between distributions. If mass  functions have similar shapes, 
trends have to be similar. Flat trends mean agreement with the shape of the low-$z$ observed field mass function.  

Discrepancies between observations and sim-projections are evident, as already discussed in the previous section.

In the observed samples, at a given redshift cluster and field trends are similar. This indicates that the environment does not strongly affect the shape of the mass distributions, at both redshifts considered, at least above \M=10.25, our most conservative mass limit (see also C13 and V13). At masses larger than \M=11.4, there are differences among the different samples but error bars are too large to draw statistically robust conclusions.
The same main result is found when considering 
 the sim-projected samples. 
 Hence, sim-projections, though not reproducing all the environments separately,  can reproduce the observed invariance of the shape of the mass function with the environment above \M=10.25. 
We note that, in the low-$z$ clusters, the exclusion of galaxies  less massive than \M=10.25 reduces the discrepancies between the observed and sim-projected mass functions, making possible this result. This highlights the  importance of reaching a broad mass dynamical range to robustly determine
the shape of the galaxy mass function.
We do not have observational data at lower masses for galaxies in all the environments, however, sim-projections predict similar mass functions for galaxies in different environments, at a given redshift.

G11, exploiting the increased resolution provided by the Millennium-II Simulation \citep{bk09} showed  that in their simulations the stellar mass function of galaxies in
rich clusters at low-$z$ is predicted to be very similar in shape to that
in the general field, even down to \M$\sim$7.

As far as the evolution is concerned, 
in observations the number density of cluster galaxies increases at increasing redshift, although not systematically as a function of stellar mass (the blue solid line is steeper than the red line). In the field, the opposite is found: the number density of galaxies tends to decrease with increasing redshift, with the decrease being more significant at lower stellar masses. 
However, 
at both redshifts, the relative slopes at low masses of clusters and field are similar, highlighting a similar change of the shape of the  mass function with redshift.\footnote{This agrees with the results presented in \cite{morph}, V13, where the authors focused on the shape of the mass function and didn't discuss the evolution of the number density.}

For the sim-projected samples, both environments exhibit an increase in the number density of galaxies with redshift. The evolution does not depend strongly on the stellar mass (i.e. the lines are shifted vertically by about the same factor as a function of stellar mass); hence the shape of the mass functions stays almost constant over the cosmic time, and the evolution
seen in observations is not reproduced. 

\section{Results: The simulated mass functions}
\begin{figure*}
\centering
\includegraphics[scale=0.44]{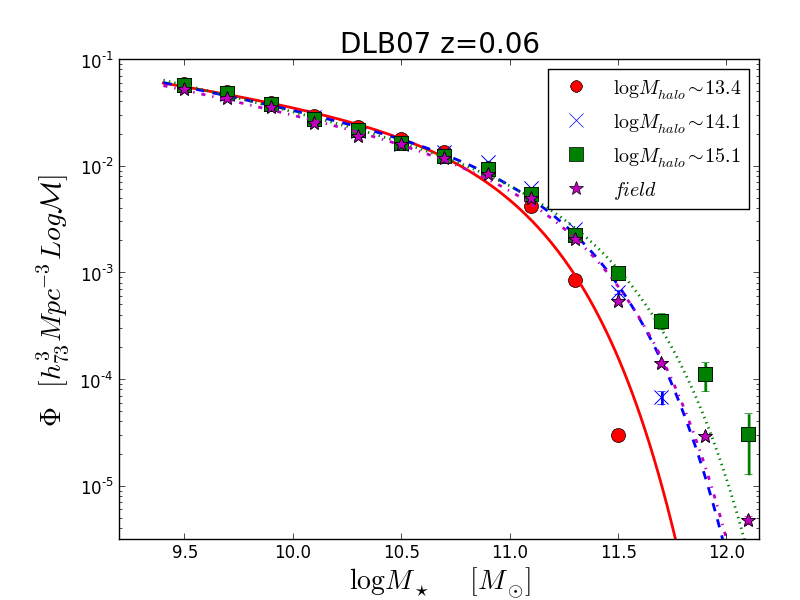}
\includegraphics[scale=0.44]{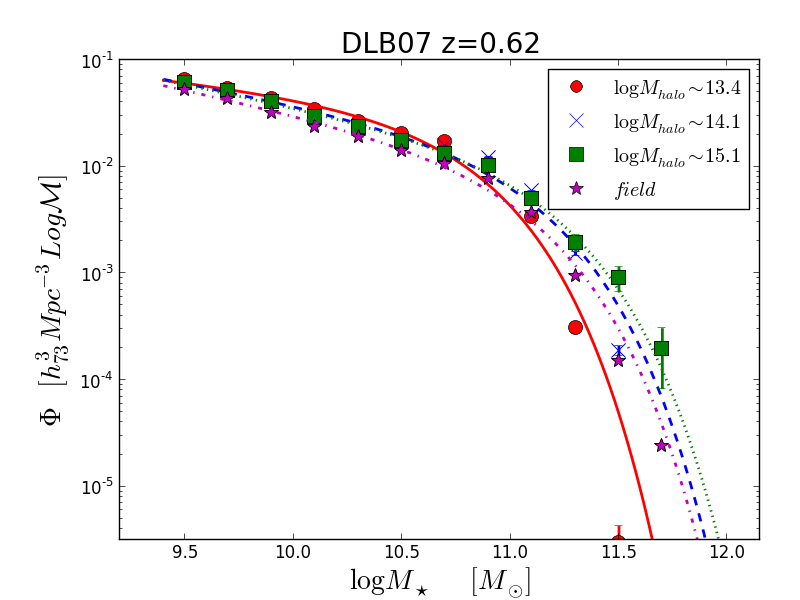}
\includegraphics[scale=0.44]{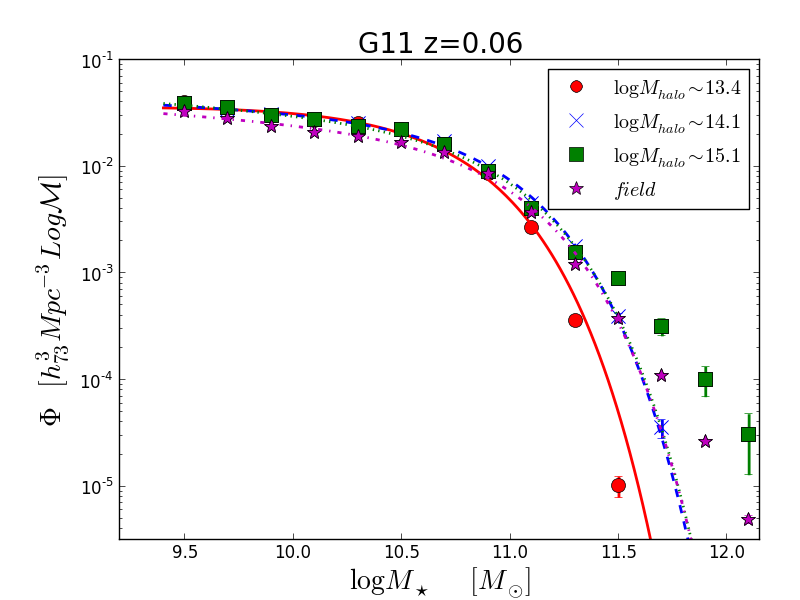}
\includegraphics[scale=0.44]{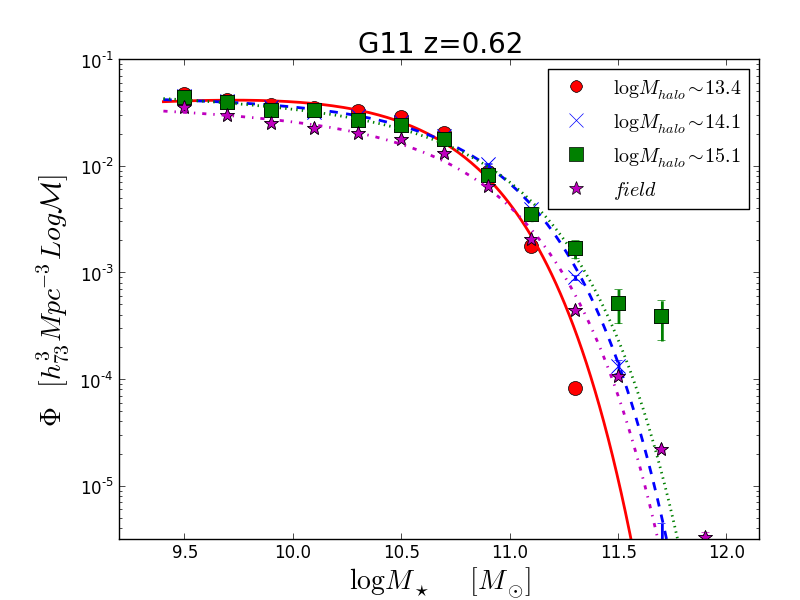}
\caption{Simulated mass function and Schehcter fits for galaxies in structures with different $M_{halo}$, within $1R_{200}$   (red points and solid line: galaxies in structures with $\log M_{halo}\sim 13.4$, blue crosses and dashed line: galaxies in structures with $\log M_{halo}\sim 14.1$, green squares and dotted line: galaxies in structures with $\log M_{halo}\sim 15.1$) and field galaxies (purple stars and dashed-dotted line), at z=0.06 (left panels) and z=0.62 (right panels), for the DLB07 model (upper panels) and the G11 model (bottom panels).   Normalization, errorbars on the $y-$axis, lines, and labels 
are as in Fig. \ref{cfr_simobs_obs}. Given the uncertainties on the best fit parameters are very 
small, we do not show error contours.
The typical errorbar on the $x-$axis is 0.1 $dex$, corresponding to the width of the bin \label{sim_Mhalo}}
\end{figure*}

In the previous section we considered observations and sim-projections, finding that in the local universe the semi-analytic model well reproduces the general field and its finer environments, while it fails in reproducing the cluster mass function. So there appears to be a break between groups and clusters. Observationally, groups and clusters have similar mass functions (C13), in sim-projections they don't (plot not shown). In this section we try to analyze more carefully the transition between groups and clusters, exploiting only simulations. In this way, we can control the mass of the halos, and inspect whether and at which halo mass differences emerge. 
We consider three different halo mass bins: halos of $\log M_{halo}\sim 13.4$ ({\it least massive haloes}) roughly correspond to groups, halos of  $\log M_{halo}\sim 14.1$ ({\it intermediate massive haloes}) to intermediate-mass clusters, halos of $\log M_{halo}\sim 15.1$  ({\it most massive haloes})
to massive clusters. 
Using only simulations, we can also extend to larger clustercentric distance and inspect whether galaxies at different distances are regulated by different mass distributions.\footnote{An analysis of this kind is not possible with our observations, since we do not have information yet on the external regions of these structures. }
For completeness, we present a similar analysis also at higher redshift. 
We take advantage of two different semi-analytic models to inspect if their different recipes produce different results. We consider the 
DLB07 and G11 models
and the samples presented in \S\ref{sim_sample} and consider all galaxies more massive than \M=9.4.

In Figure \ref{sim_Mhalo} we show how the mass function changes across different environments, from the field to massive halos, separately for the two models. We remind the reader that the field include  galaxies in {\it all} halos above the resolution of the Millennium at that redshift. 

For \M$<11$, in both models and at both redshifts, mass functions in  different environments are almost indistinguishable. 
The DLB07 and G11 models give different predictions  for the shape of the mass function: the former predicts a steeper mass function than the latter, which flattens out at \M$\sim10.6$ in all the environments (see also Table \ref{tab:fit_sim}). In addition, with the adopted normalization, the G11 field shows a systematically lower number density than the other environments considered, in particular at high-$z$. The same is true for the field at high-$z$ in the DLB07 model, although the effect is less significant. 

The different shape of the mass function in the two models is mainly due to the stronger stellar feedback adopted in the G11 model, that, together with the introduction of a model for stellar stripping, contributes to reduce the number of intermediate to low-mass galaxies with respect to the DLB07 model. 

Mass distributions differentiate with environment at high stellar masses:
the maximum mass reached by galaxies in the most massive halos is higher than that reached by galaxies in lower-mass halos. This is more noticeable at low-$z$, where, in both models, the maximum mass varies from \M=11.5 in the least massive systems to \M=12.1 in the most massive ones. At higher redshift, the difference is only 0.2 $dex$. 
The field mass function reaches a mass of \M=12.5 at low redshift in both models, and 12.1 (DLB07) and 12.3 (G11) at higher redshift. 

The number density of the most massive galaxies increases with the mass of the halos. 
This is particularly evident in the G11 model and at low redshift where the mass function of halos with $\log M_{halo}\sim 15.1$  exhibits a strong excess of massive galaxies and deviates significantly from the exponential shape provided by the best fit Schechter function. 
 In this case the analytical fits  (Table \ref{tab:fit_sim}), do not fully describe the measured mass distributions. 
 
\begin{figure*}
\centering
\includegraphics[scale=0.44]{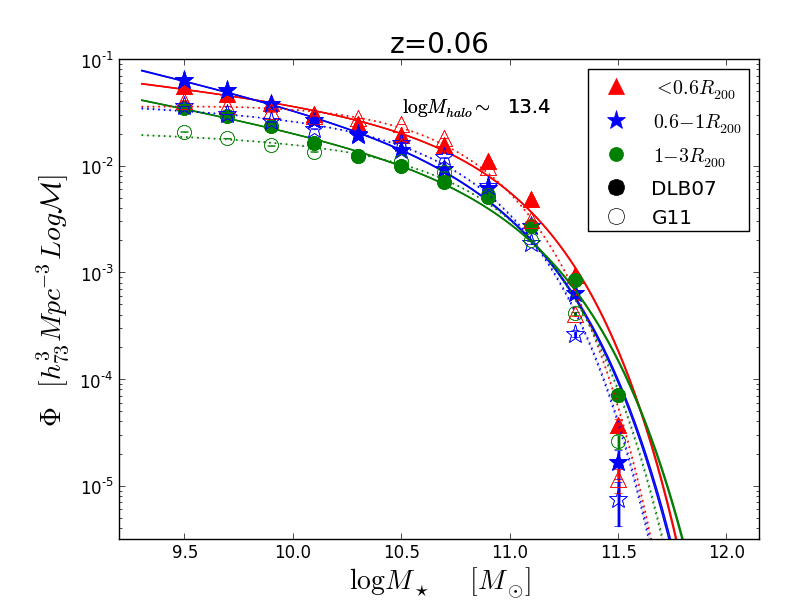}
\includegraphics[scale=0.44]{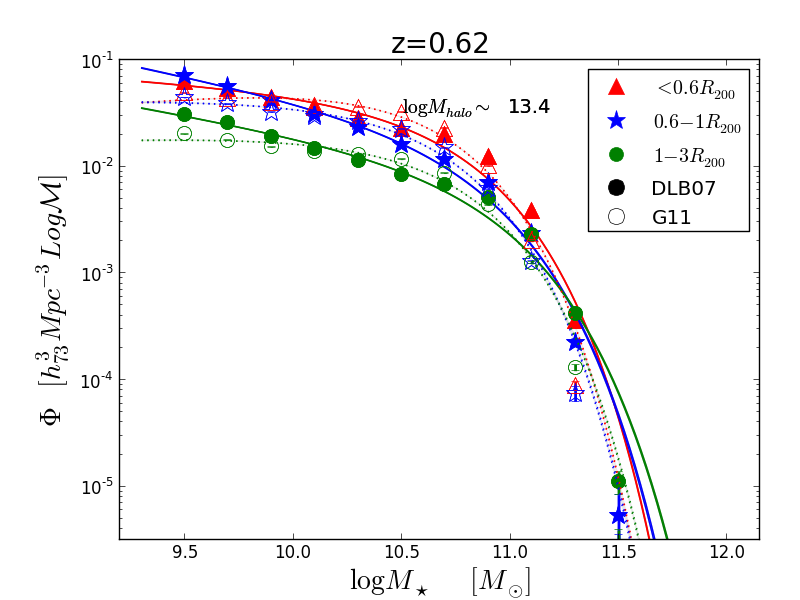}
\includegraphics[scale=0.44]{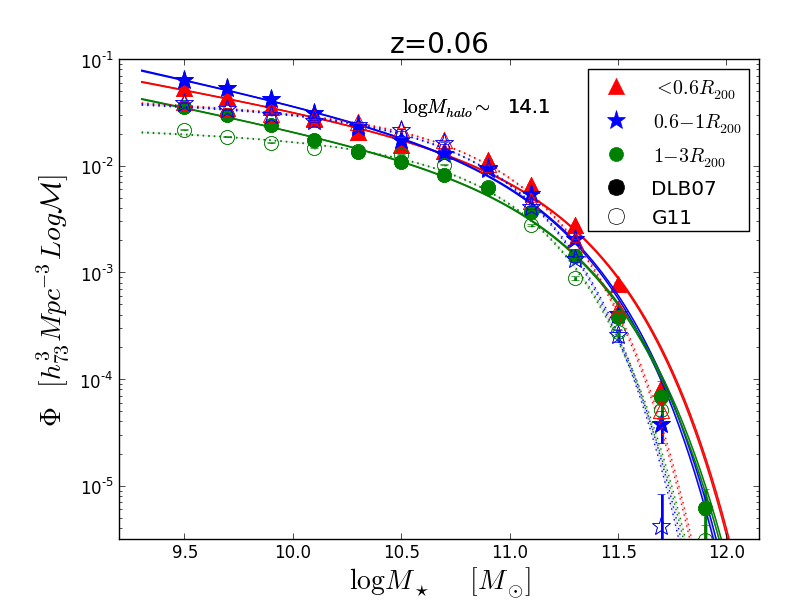}
\includegraphics[scale=0.44]{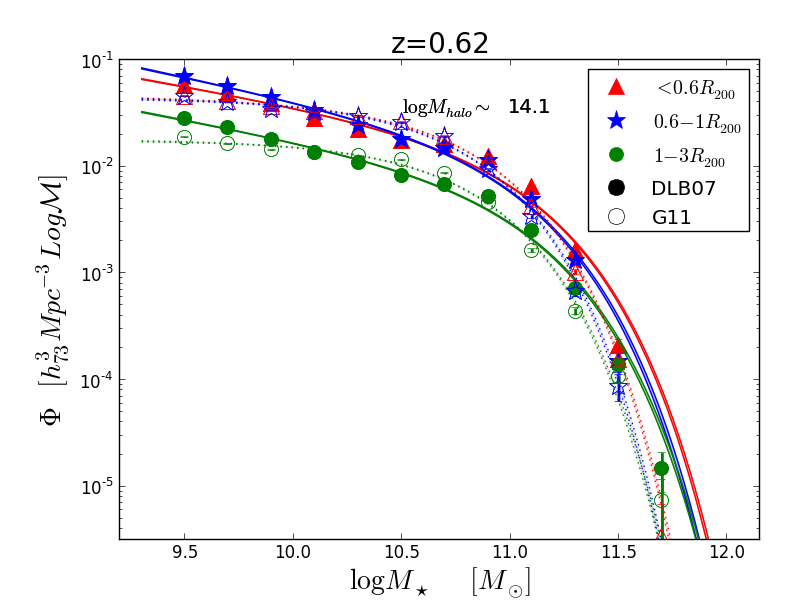}
\includegraphics[scale=0.44]{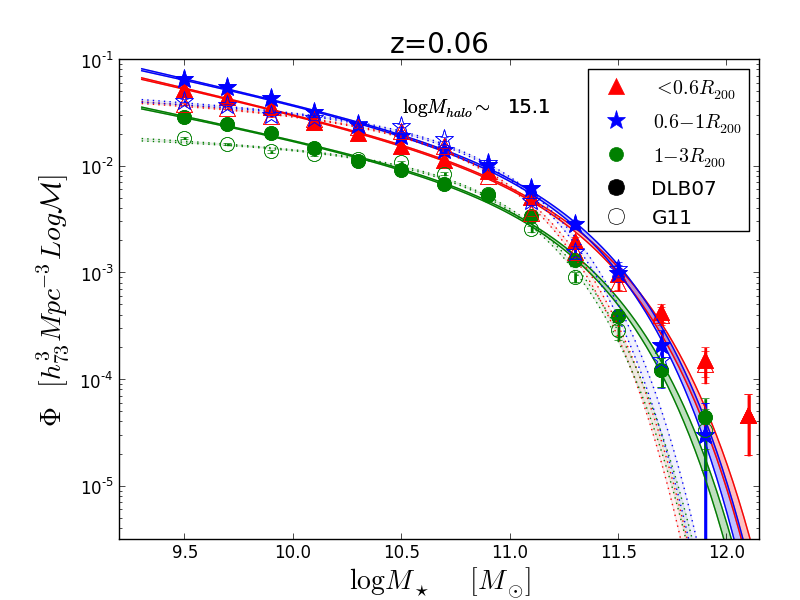}
\includegraphics[scale=0.44]{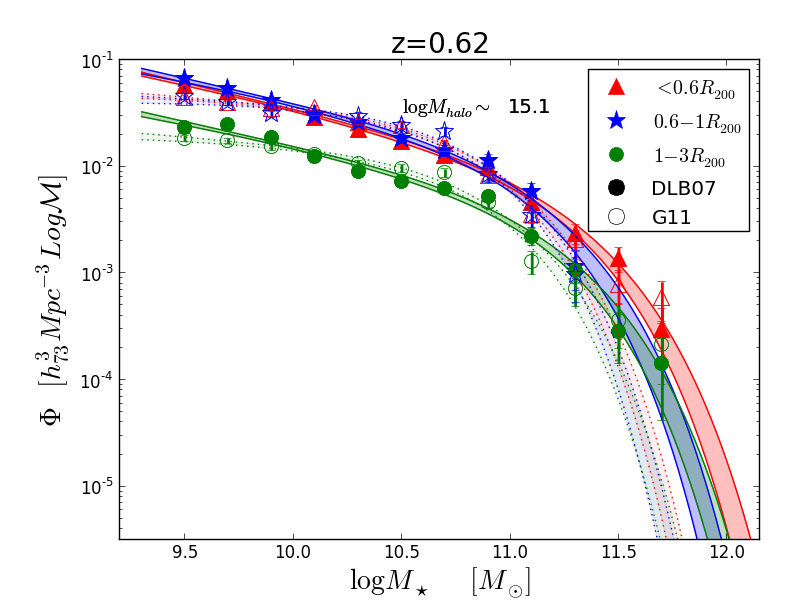}
\caption{Simulated mass function of galaxies in structures of the same halo mass, but at different cluster centric distances, at low (left panels) and intermediate-$z$ (right panels), for the DLB07 (solid symbols and darker shaded areas) and the G11 (empty symbols and lighter shaded areas) models. The mass of the halos is indicated in the upper part of each panel. Red symbols represent galaxies within $0.6 R_{200}$, blue symbols galaxies within 0.6-1 $R_{200}$, green symbols galaxies within $1-3 R_{200}$. Normalization, errorbars on the $y-$axis, lines, shaded areas and labels are as in Fig. \ref{cfr_simobs_obs}.  Given the uncertainties on the best fit parameters are very 
small, we do not show error contours.
\label{sim_r200}}
\end{figure*}

In general, Schechter fits
show a variation of the parameters from the least massive environment - characterized by larger $\alpha$ and smaller $M^*$ - to the most massive one - characterized by smaller $\alpha$ and larger $M^*$. 
The best-fit parameters describing the mass functions are always statistically different at 3-$\sigma$ level, except for the field, intermediate massive and most massive halos in the G11 model, where differences are significant only at $1-\sigma$ level (see  Table \ref{tab:fit_sim}).

In both models and at all redshifts, the field mass function is not the steepest one: it resembles the one of intermediate massive halos. 
This may be the result of the fact that these particular fields are dominated by halos of this mass.\footnote{Given the resolution limit of the simulations, halos of $\log M_{halo}\sim 12$ are barely resolved in the simulations.  
If one considers a higher resolution simulation, the result might change.}

Hence, in the simulated samples the environment is able  to  influence both the shape of the mass function and the maximum mass.

\subsection{Halo-centric distance}
We now 
focus only on galaxies in halos and  analyze if, at any halo mass, simulations predict a dependence of the mass distribution on   clustercentric distance. Some change is expected because going from the center toward the outskirts, galaxies  suffer different processes.  The external parts of halos can be seen as  intermediate and transitional regions between the cores and the field. They contain both galaxies just fallen in the structure and galaxies that, lying on very eccentric orbits, might have experienced more than one passage through the cluster core region.

In Figure \ref{sim_r200}
we show the mass function of galaxies within $0.6R_{200}$, within $0.6-1 R_{200}$ and  within $1-3 R_{200}$, at low (left panels) and intermediate-$z$ (right panels) respectively, for halos of different mass and for both models.

In the least massive systems (upper panels), and at both redshifts considered,  both the DLB07 and G11 mass functions seem to slightly depend on the halo centric distance: the innermost regions ($r<0.6R_{200}$) are characterized by flatter mass functions at  intermediate masses (10$<$\M$<$11) with respect to the other regions. In addition, galaxies at 1$<r/R_{200}<$3 always show a lower number density, but this might be due to the adopted normalization (see Appendix \ref{app_normalization}). 
The parameters of the Schechter fit (see Table \ref{tab:fit_sim})  
support these discrepancies. The mass range spanned by galaxies at the different halo centric distances is the same. 
Comparing the two models, at high an low masses the G11 model  predicts a smaller number of galaxies than the DLB07 one, instead at intermediate masses it predicts a slightly higher number density, in all environments.  The consequence  is that the DLB07 mass functions are  steeper. 

In halos of intermediate mass (central panels), at both redshifts and for both models separately, the  shapes of the mass function seem very slightly  dependent on clustercentric distance, as also supported by the Schechter fit. Intermediate regions show an excess of low-mass galaxies, making the mass function steeper. Only the external regions are systematically characterized by a smaller number density. This result is qualitatively in agreement with the observational results: V13, exploiting EDisCS data ($M_{halo}\sim 14$),  showed how galaxies within and outside the virialized regions of clusters  have  similar mass function. 
Both models predict that very massive galaxies (\M$\sim$11.9 at low-$z$, \M$\sim$11.7 at higher) might be found only in the outer regions. This result might be surprising, however, similar findings have been found by  \cite{delucia04a}, who analyzed the radial distribution of substructures. The radial profile of substructure number density is ``antibiased'' relative to the dark matter profile in the inner regions of halos. The most massive substructures reside preferentially in the outer regions of halos. They suggest that this might be due  to the fact that substructures undergo substantial tidal stripping in the dense inner regions of halos.

In the most massive halos (bottom panels) mass function shapes are similar up to \M$\sim$11, then they differentiate: there is an excess of very massive galaxies in the halo cores. Both the excess and the maximum masses are similar in the two models: \M$\sim$12.1 at low-$z$, \M$\sim$11.7 at high-$z$. 
Best fit parameters are in agreement at $1-2-\sigma$ level, probing on an analytical ground the similarities of these mass functions. 

To summarize, the DLB07 and G11 models qualitatively lead to similar results. However, 
the mass functions obtained from the two models are different. In the low mass regime, 1) the G11 mass functions is flatter than the DLB07 one,  
2) the G11 model predicts similar mass functions for galaxies within $0.6R_{200}$ and within $0.6-1 R_{200}$, while the DLB07 model shows an excess of low-mass galaxies in the outer regions, in the least massive and intermediate massive systems. 
Finally, the  shape of the mass function does not strongly depend on the halo centric distance, 
at fixed redshift and halo mass. However, some differences are found for galaxies in the cores of the least massive and most massive systems considered.

\subsection{Evolution in different environments}
The last step is to test whether simulations predict an evolution for galaxies in the different environments. For galaxies in halos, we consider only galaxies within the virial radius.

\begin{figure*}[!t]
\centering
\includegraphics[scale=0.44]{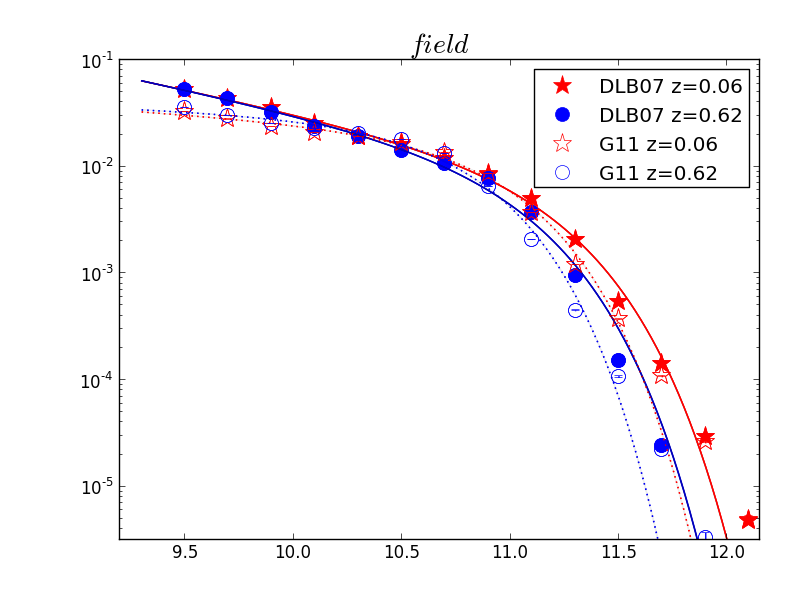}
\includegraphics[scale=0.44]{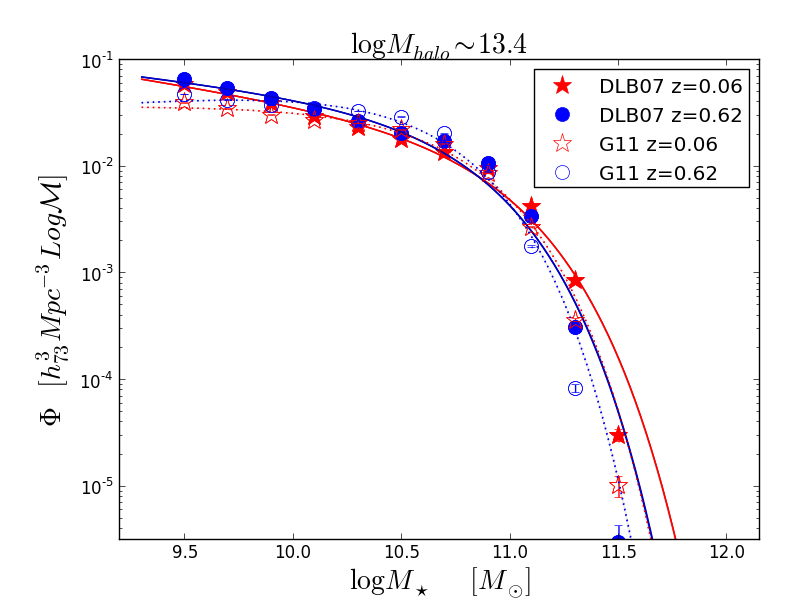}
\includegraphics[scale=0.44]{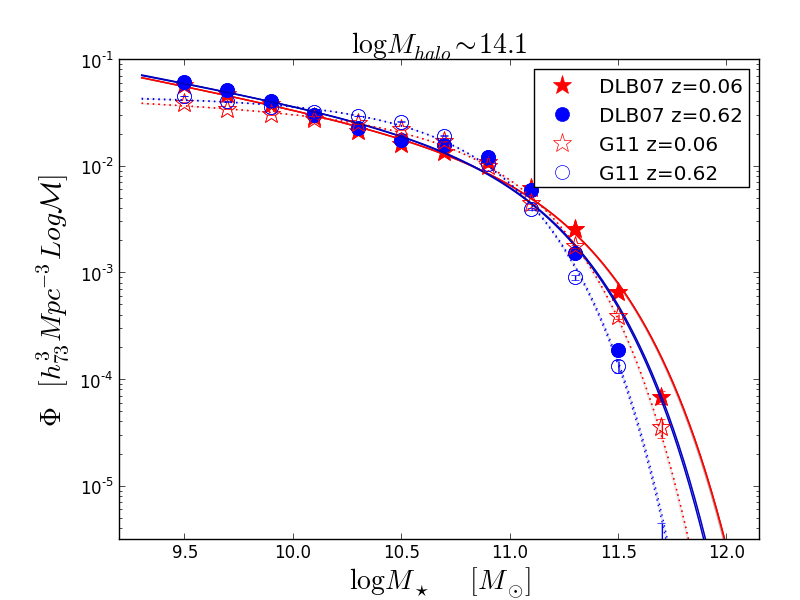}
\includegraphics[scale=0.44]{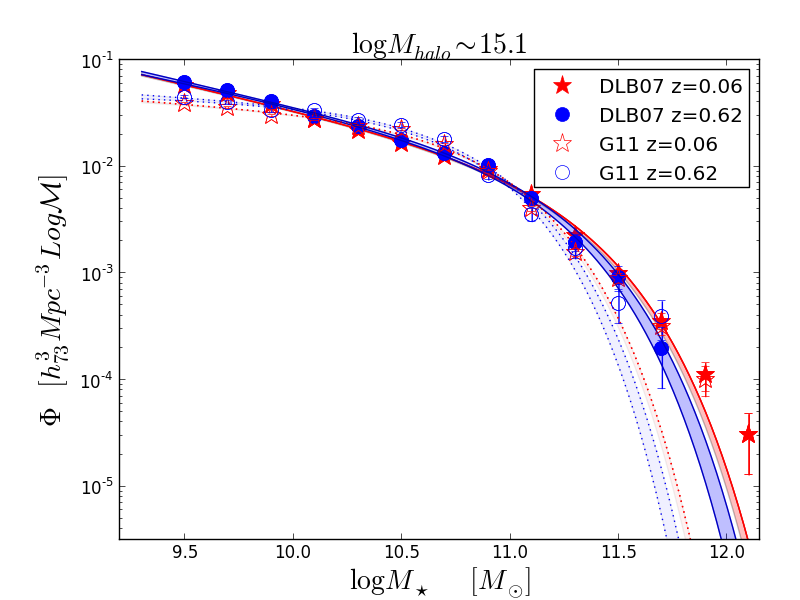}
\caption{Evolution of the simulated mass function in each environment, for the DLB07 (solid symbols and darker shaded areas) and G11 (empty symbols and lighter shaded areas) models. The environment is indicated in the upper part of each panel. Red symbols represent galaxies at $z=0.06$, blue symbols  galaxies at $z=0.62$. Normalization, errorbars on the $y-$axis, lines, shaded areas and labels are as in Fig. \ref{cfr_simobs_obs}.   Given the uncertainties on the best fit parameters are very 
small, we do not show error contours.
\label{cfr_sim_evol}}
\end{figure*}

In the field (upper left panel of Fig. \ref{cfr_sim_evol}), both models predict an evolution of the high mass end of the mass functions. In  the local universe galaxies with \M$>11.2$ are more numerous than in the distant one. Also the maximum mass evolves, increasing of $\sim0.4\, dex$ in the DLB07 model, $\sim0.2\, dex$ in the G11 one. In contrast, at low masses no significant evolution is detected.  
An analogous evolution in the high mass end of the mass function - and consequently in $M^*$ -  is detected 
in halos in all mass bins considered, though this is less significant probably due to the lower number statistics.

In the lowest halo mass bin considered, no evolution of the maximum mass is detected. In both models and at both redshifts there are no galaxies more massive than \M$\sim$11.5. 
The evolution of the massive end is somewhat more important in the G11 model. 

In the intermediate mass halos, the evolution of the shape of the mass function is very similar in the two models, 
but a significant evolution of the maximum mass (from \M$\sim$11.5 to $\sim$11.9) is found only in the DLB07 model.

In the most massive halos, differences between mass functions emerge only for  \M$>11.5$, In these systems,  also the maximum mass depends on redshift: in the distant universe there are no galaxies more massive than \M=11.7, while in the local universe galaxies with \M$\sim$12.1 are found. 

Where the fits  describe well the mass functions, they support the results, showing  a quite strong evolution of $M^*$, but a very little evolution of $\alpha$. As already mentioned, this is not the case for the most massive halos. 

To conclude,  the shape of the massive end of the mass function  evolve, in all environments considered. In contrast, no evolution is found for low-mass galaxies.

\begin{table*}
\caption{Simulated sample: Best-fit Schechter function parameters ($M_\star^{*}$, $\alpha$,  $\Phi^{*}$) for the mass functions of galaxies in different environments down to \M=9.4. 
\label{tab:fit_sim}}
\centering
\begin{tabular}{ll|ccc|ccc}
	\multicolumn{8}{c}{{\bf DLB07}}		 \\
	&& \multicolumn{3}{c|}{$z=0.06$}		&	\multicolumn{3}{c}{$z=0.62$}			 \\
\hline
								&&$\log (M_\star^*/M_\odot) $ & 	$\alpha$ &$ \Phi^{*} $	 & $\log (M_\star^*/M_\odot) $ & 	$\alpha$ &$ \Phi^{*} $	\\
\multicolumn{2}{l|}{$M_{halo}\sim 13.4$}	& 	&   	& \\
	&$R/R_{200}<0.6$	& 10.832$\pm$0.002	&-1.233$\pm$0.003  	&0.01173$\pm$0.00009 	&10.672$\pm$0.002		&-1.127$\pm$0.003   	&0.01890$\pm$0.001 \\
	&0.6<$R/R_{200}<1$	& 10.857$\pm$0.005	&-1.472$\pm$0.004  	&0.0065$\pm$0.0001	&10.758$\pm$0.001		&-1.407$\pm$0.004   	&0.0096$\pm$0.0001 \\
	&$R/R_{200}<1$		& 10.848$\pm$0.002	&-1.314$\pm$0.002  	&0.00960$\pm$0.00006 	& 10.708$\pm$0.002	&-1.218$\pm$0.002  	&0.01523$\pm$0.00008 \\	
	&$1<R/R_{200}<3$	& 10.936$\pm$0.003 	&-1.393$\pm$0.003  	&0.00424$\pm$0.00005  &10.858$\pm$0.003		&-1.355$\pm$0.003  	&0.00444$\pm$0.00005  \\
\multicolumn{2}{l|}{$M_{halo}\sim 14.1$}	& 	&   	&  \\
	&$R/R_{200}<0.6$	& 11.13$\pm$0.05	&-1.38$\pm$0.04  	&0.0056$\pm$0.0001 	& 11.004$\pm$0.006	&-1.344$\pm$0.005   	&0.0075$\pm$0.0001 \\
	&0.6<$R/R_{200}<1$	& 11.07$\pm$0.08	&-1.44$\pm$0.06  	&0.0059$\pm$0.0001 	& 10.97$\pm$0.01	&-1.412$\pm$0.008   	&0.0075$\pm$0.0002 \\
	&$R/R_{200}<1$		& 11.10$\pm$0.04	&-1.397$\pm$0.003  	&0.0058$\pm$0.0001  & 11.008$\pm$0.005	&-1.380$\pm$0.004   	&0.0071$\pm$0.0001 \\
	&$1<R/R_{200}<3$	& 11.12$\pm$0.07	&-1.40$\pm$0.05 	&0.0034$\pm$0.0001 	& 11.017$\pm$0.008	&-1.376$\pm$0.007   	&0.0032$\pm$0.0001  \\
\multicolumn{2}{l|}{$M_{halo}\sim 15.1$}	& 	&   	&  \\
	&$R/R_{200}<0.6$	& 11.19$\pm$0.02	&-1.45$\pm$0.01   	&0.0041$\pm$0.0002 	& 11.16$\pm$0.06	&-1.43$\pm$0.03		&0.0050$\pm$0.0007\\
	&0.6<$R/R_{200}<1$	& 11.17$\pm$0.03	&-1.43$\pm$0.01   	&0.0054$\pm$0.0003 	& 11.17$\pm$0.07	&-1.44$\pm$0.04		&0.005$\pm$0.001\\
	&$R/R_{200}<1$		& 11.22$\pm$0.02	&-1.47$\pm$0.01		&0.0045$\pm$0.0007	& 11.08$\pm$0.04	&-1.40$\pm$0.02   	&0.0061$\pm$0.0006   \\
	&$1<R/R_{200}<3$	& 11.18$\pm$0.02	&-1.41$\pm$0.01	 	&0.0040$\pm$0.0002 	& 11.18$\pm$0.06	&-1.41$\pm$0.04		&0.0021$\pm$0.0004\\
    &					&					&					&					&					&					&					\\
\multicolumn{2}{l|}{$field$} 	& 11.13$\pm$0.01	&-1.42$\pm$0.01  	&0.0047$\pm$0.0003 & 10.99$\pm$0.01	&-1.43$\pm$0.01 	&0.0053$\pm$0.0003  \\
	\\
	\multicolumn{8}{c}{{\bf G11}}		 \\
	&& \multicolumn{3}{c|}{$z=0.06$}		&	\multicolumn{3}{c}{$z=0.62$}			 \\
\hline
								&&$\log (M_\star^*/M_\odot) $ & 	$\alpha$ &$ \Phi^{*} $	 & $\log (M_\star^*/M_\odot) $ & 	$\alpha$ &$ \Phi^{*} $	\\
\multicolumn{2}{l|}{$M_{halo}\sim 13.4$}	& 	&   	& \\
	&$R/R_{200}<0.6$	& 10.658$\pm$0.003	&-0.931$\pm$0.005  	&0.0207$\pm$0.0001 	&10.534$\pm$0.004		&-0.80$\pm$0.01   	&0.0324$\pm$0.0005 \\
	&0.6<$R/R_{200}<1$	& 10.685$\pm$0.004	&-1.069$\pm$0.005  	&0.0125$\pm$0.0001	&10.560$\pm$0.003		&-0.965$\pm$0.005   	&0.0204$\pm$0.0002 \\
	&$R/R_{200}<1$		& 10.667$\pm$0.002	&-0.972$\pm$0.004  	&0.0177$\pm$0.0001  &10.544$\pm$0.007		&-0.849$\pm$0.01  	&0.0282$\pm$0.0006 \\
	&$1<R/R_{200}<3$	& 10.774$\pm$0.003 	&-1.053$\pm$0.004   &0.00745$\pm$0.00007 	&10.633$\pm$0.005		&-0.926$\pm$0.007  	&0.0099$\pm$0.0001  \\
\multicolumn{2}{l|}{$M_{halo}\sim 14.1$}	& 	&   	&  \\
	&$R/R_{200}<0.6$	& 10.885$\pm$0.006	&-1.091$\pm$0.005  	&0.0127$\pm$0.0001 	& 10.771$\pm$0.005	&-1.036$\pm$0.007   	&0.0175$\pm$0.0003 \\
	&0.6<$R/R_{200}<1$	& 10.822$\pm$0.007	&-1.061$\pm$0.007  	&0.0135$\pm$0.0002 	& 10.720$\pm$0.007	&-1.03$\pm$0.01   	&0.0181$\pm$0.0004 \\
	&$R/R_{200}<1$		& 10.865$\pm$0.004	&-1.081$\pm$0.004  	&0.0128$\pm$0.0005  & 10.745$\pm$0.004	&-1.018$\pm$0.005   	&0.0182$\pm$0.0002 \\
	&$1<R/R_{200}<3$	& 10.871$\pm$0.006	&-1.066$\pm$0.006 	&0.0073$\pm$0.0001 	& 10.757$\pm$0.008	&-1.000$\pm$0.009  	&0.007$\pm$0.0002  \\
\multicolumn{2}{l|}{$M_{halo}\sim 15.1$}	& 	&   	&  \\
	&$R/R_{200}<0.6$	& 10.84$\pm$0.02	&-1.12$\pm$0.02  	&0.0116$\pm$0.0005 	& 10.76$\pm$0.04		&-1.04$\pm$0.04   	&0.017$\pm$0.002 \\
	&0.6<$R/R_{200}<1$	& 10.91$\pm$0.02	&-1.13$\pm$0.02  	&0.014$\pm$0.0007 	& 10.86$\pm$0.05		&-1.11$\pm$0.06   	&0.013$\pm$0.002 \\
	&$R/R_{200}<1$		& 10.89$\pm$0.01	&-1.13$\pm$0.01   	&0.0111$\pm$0.0004 	& 10.81$\pm$0.03		&-1.08$\pm$0.03   	&0.014$\pm$0.001   \\
	&$1<R/R_{200}<3$	& 10.90$\pm$0.02	&-1.08$\pm$0.02  	&0.0061$\pm$0.0003 	& 10.75$\pm$0.05		&-1.01$\pm$0.05 		&0.0074$\pm$0.0008  \\
    &					&					&					&					&					&					&					\\
\multicolumn{2}{l|}{$field$} 	& 10.90$\pm$0.02	&-1.127$\pm$0.002  	&0.0091$\pm$0.00005 & 10.729$\pm$0.003	&-1.070$\pm$0.004 	&0.0122$\pm$0.0001  \\
\end{tabular}
\end{table*}

\section{Discussion} 
\subsection{Observations and sim-projections}
In the first part of the paper we have tested whether simulations are able to reproduce the observational results for the galaxy stellar mass function in different environments, at $z\sim0$ and $z\sim0.6$.

Being able to match the field (and its finer environments) mass function in the local universe down to the completeness limit of the observational sample (\M$>10.25$), the DLB07 model fails in reproducing the observed mass function of field galaxies at higher $z$ and that of clusters at both epochs. In all three cases, the sim-projected mass function is steeper than the observed one in the low mass regime (the mass at which it occurs is different in the different environments and at the different redshifts), indicating an excess of low-mass galaxies.  As far as the high mass end is concerned, sim-projections predict a slightly higher number density  in clusters at low-$z$ and a slightly lower number density in both environments at higher-$z$ than those observed.
We stress that in this mass regime uncertainties on estimates are large, hence comparisons have to be interpreted with caution. 


Our results for the field are in line 
with several studies. 
As already mentioned in \S 1,  e.g. G11 and
\cite{bower12} have shown that
the semi-analytic models can be tuned to reproduce well the $z = 0 $ mass function. 
At higher redshift, e.g.,  \cite{fontana06,marchesini09, drory09}; F09; \cite{pozzetti10} and  G11
have already largely discussed the existing tension between simulations and observations
and showed that semi-analytic models over-produce the number density of low and intermediate mass functions. Indeed they predict a different shape at any mass and redshift explored.

In the studies above, observed and simulated stellar masses have been computed in different ways. Hence, even though F09 pointed out that the disagreement can be only partially resolved by the bias introduced by uncertainties in the mass determination, the comparisons might have been influenced by the different methods followed. In a recent paper, \cite{croton13} discussed the riskiness of comparing quantities obtained using different cosmologies and how delicate is the conversion between different $h$ values (see the paper for details). In our approach, we compute stellar masses in the same way for sim-projections and  observations, so any discrepancies due to the method are better under control.

The cluster environment has not been carefully investigated yet using semi-analytic models.
\cite{weinmann11b}, using the G11 model, 
compared some properties of galaxies in  nearby
clusters, without considering stellar masses. 
 They
found that  abundances, velocity dispersions and
number density profiles are reproduced well by the model. 
However simulated clusters could reproduce the red fraction of galaxies only for 
Coma and Perseus, but not for Virgo. 
Simulations have also been found to predict a too high red fraction 
compared to observations, and the authors  argued that this is likely mainly due to
an overestimate of environmental effects in the model, possibly related to overefficient ram pressure stripping of the extended gas reservoir of group galaxies.  This problem may
be exacerbated by insufficient tidal disruption of low mass
galaxies in the semi-analytic  model, which lets too many old, red dwarf
galaxies survive.

Our findings too suggest that environmental effects are not well reproduced and are in line with the results presented by \cite{contini13}, who showed that discrepancies might be alleviated including a model for stellar stripping, but they are  not solved. 

The inability of  semi-analytic models in reproducing the cluster mass function in the local universe indicates that the problem can not be limited at an over efficient star formation only at early epochs (see also \citealt{wang12}). 
It suggests that environmental effects are incorrectly included in models and the  evolution of the satellites might be not properly accounted for (see also \citealt{weinmann06b}).

Previous studies did not compare any cluster property at higher redshift. Our results 
show how the overestimate of low-mass galaxies  is even more evident than in the local universe.

\subsection{Satellites and orphans}
\begin{figure}
\centering
\includegraphics[scale=0.44]{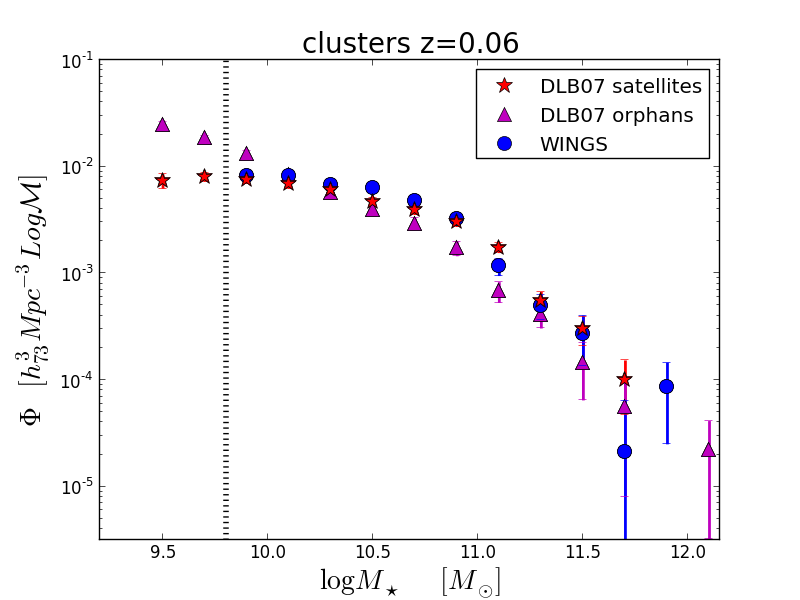}
\caption{Stellar mass function for low-$z$ cluster satellite (red stars) and orphan (magenta triangles) galaxies separately. The WINGS mass function (blue points) is reported for comparison.  Normalization, errorbars, and lines are as in Fig. \ref{cfr_simobs_obs}.
\label{orphan}}
\end{figure}
We note that in clusters we  consider all non central galaxies as satellites, even though, 
as explained in \cite{springel01, delucia04a}, models make a distinction between satellites and orphans. The formers are galaxies attached to dark matter substructures and they were previously the central galaxy of a halo that merged to form the larger system in which they currently reside. The latters are galaxies no longer associated with distinct dark matter substructures, and their stellar mass is assumed not to be affected by the tidal stripping that reduces the mass of their parent halos. They may later merge into the central galaxy of their halo.

Figure \ref{Mmax_sigma} showed that the model reproduces the observed $M_\ast -\sigma$ relation also for satellite+orphan galaxies. A more accurate analysis showed that half of the most massive galaxies are orphans, half satellites.  On the other hand, Figure \ref{cfr_simobs_obs} showed that the model does not reproduce the mass function for the same galaxies.
Indeed, inspecting separately the mass distribution of  orphans and satellites (Fig. \ref{orphan}) we find that they are characterized by different mass distributions.
Orphan galaxies are characterized by a very steep mass function and they dominate at low masses, while satellite galaxies show a mass function very similar to the observed one (see also \citealt{saro08}). 
It has been shown that the presence of orphan galaxies is fundamental to reproduce well several properties,  i.e. the clustering of the structures and the differences between the galaxy and subhalo profiles in the inner regions of  clusters \citep{gao04, wang06}. However, the fact that excluding orphans reduces the discrepancies between the observed and sim-projected mass functions is a tantalizing result that could suggest the treatment of orphan galaxies might be improved to help solving the problem.

\subsection{The reasons for the discrepancies}
The excess of low-mass galaxies at high-$z$ in the models might be due to the fact that these galaxies are predicted to form too early and have too little ongoing star formation at later times (e.g. F09, \citealt{firmani10}, G11, \citealt{weinmann12}).
The same discrepancy has been reported also by \cite{weinmann12} in two
state-of-the-art cosmological hydrodynamical simulations, highlighting that the
problem is fundamental.
Indeed, they showed that
a key problem is the presence of a positive instead of a negative correlation between specific
star formation rate and stellar mass.
A similar correlation characterizes also the specific dark matter halo accretion rate and the halo mass, indicating that in the models
the growth of galaxies follows the growth of their host halos too closely.
Therefore, it is necessary to find a mechanism that decouples the growth of low
mass galaxies, which occurs primarily at late times, from the growth of their host
halos, which occurs primarily at early times. 
\cite{weinmann12} then argued that the current form of star formation driven feedback implemented in most galaxy formation models is unlikely
to achieve this goal, owing to its fundamental dependence on host halo mass and time.

It has also been suggested that  AGN feedback could provide a solution to
the ``downsizing problem'' \citep{bower06, croton06}. 
Indeed, the suppression of late gas condensation in massive halos causes shorter formation time-scales for more massive galaxies \citep{delucia06}, in qualitative agreement with the observed
trends. However, as discussed also by F09,  \cite{somerville08} showed that the predicted trends may not be as strong as the observed
ones, even in the presence of AGN feedback.

F09 showed that the main contributors at all redshifts to the low-mass end excess are low-mass ($10^{11} < M_h/M_\odot < 10^{12}$) DM halos.
Models predict a roughly constant number density of low-mass central galaxies, while the low-mass satellite population shows a gradual increase which F09 argue is due to the infall of galaxies from the surrounding areas into clusters. This implies that small objects are overproduced while they are central galaxies, and the excess is not primarily due to inaccuracies in the modeling of satellites, even though it manifests in their mass distribution.  Therefore, mechanisms that only impact satellite galaxies (such as ram pressure stripping) 
can not solve this problem.
On the contrary, it would be important to find a physical process that can suppress the star formation in central galaxies hosted by intermediate to low-mass halos ($M_h/M_\odot < 10^{12}$).

\cite{hopkins13} have recently improved the treatment of the  ISM  and stellar feedback in a series of high-resolution cosmological simulations. They included  cold molecular through atomic, ionized, and hot diffuse ISM a took the stellar feedback inputs (energy, momentum, mass, and metal fluxes)  from stellar population models, without free/adjustable parameters. These simulations lead to a reasonable concordance with the stellar mass function for \M$<11$,   suggesting that actually the effects of stellar feedback on galaxy star formation histories and stellar masses might be correctly modeled.

\subsection{Clusters and groups}\label{cl_gr}
We have shown that at low-$z$ the semi-analytic model correctly
reproduces the observed field finer divisions, groups included, at least above the observed completeness limit. In contrast, it can not reproduce the cluster mass function at similar redshift. This suggests an inaccuracy in the treatment of cluster specific processes.

Clusters and groups are environments where the decline of gas accretion in galaxies influences galaxy properties, but the specific processes that operate in them might be different.  However, in the semi-analytic models they are treated in a similar way. Motivated by the starvation scenario suggested by \cite{larson80}, all hot gas around the satellites is immediately removed upon infall. In the DLB07 model, the stripped gas is then made available for cooling to the central galaxy of the corresponding structure. This simple prescription leads to satellite galaxies that are too red (e.g. \citealt{weinmann06b, wang07}) compared to observations, suggesting that part of the hot gas should remain with the satellite galaxy. This is not a minor issue, since satellite galaxies constitute a significant fraction of the total galaxy population and since changing the prescription for gas stripping in satellites may have a considerable impact on the central galaxy population. First of all, satellites eventually merge with the central galaxies in their halo. If they can grow to higher stellar masses, central galaxies will become more massive as well. However, in simulations the  
merging history of
central galaxies seems to be modeled correctly, as also shown in Fig. \ref{Mhalo_Mcen}  where the $M_\ast -\sigma$ relation for central galaxies  is well reproduced by the model. Then, if part of the hot gas stays attached to satellites, the amount of gas available for cooling to the central galaxy is reduced. Finally, if satellite galaxies merging with the central galaxy still contain cold gas, the ensuing star burst will make the central galaxy bluer for a certain period of time, and will result in a higher final stellar mass.

Some attempts to treat environmental effects more realistically have  been made e.g. by \cite{kang08} and \cite{font08}. The formers showed that the  decrease of the efficiency of gas stripping in satellites leads to a fraction of blue central galaxies which is higher than observed, but it can be  counterbalanced  by the inclusion of an additional prescription for the disruption of satellites. The latters implemented a  model based on the hydrodynamical simulations of ram-pressure stripping by \cite{mccarthy08}, to obtain better agreement with observed environmental effects than previous semi-analytic models.
\cite{okamoto03} and \cite{lanzoni05} investigated the effect of ram-pressure stripping of the cold disc gas in satellites, concluding that it has a negligible impact on the results, because the complete stripping of the hot gas halo already makes the satellites passive.
\cite{weinmann10} used the DLB07 model letting the diffuse gas halo around satellites be stripped at the same rate at which the dark matter subhalo loses mass due to tidal effects. They found that observations at $z=0$ can be reproduced by a simple recipe in which 10-20\% of the
initial gas reservoir in the halo is stripped per Gyr. They didn't focused on higher redshift galaxies, where the dark matter stripping seems to be more efficient. The same authors also implemented  a model in which environmental effects arise solely due to ram-pressure stripping. They showed that this assumption is not exhaustive. Indeed, stripping effects in clusters for low-mass galaxies are too strong to explain the observations, while they are not strong enough to reproduce the observations for high-mass galaxies.
These discrepancies could indicate that  the DLB07 model like other models overestimate the hot gas mass in groups with masses between 10$^{12}$ and 10$^{14} h^{-1} M_\odot$, especially in the central, most dense regions of these systems \citep{bower08}. This could lead to too strong pre-processing of cluster galaxies in groups. 

\subsection{Simulations}
In the second part of this study, 
we  made use of the DLB07 and G11 semi-analytic models to investigate the dependence of the mass function on environment. We considered field galaxies and galaxies in halos with $\log M_{halo} \sim13.4$ (least massive halos), $\log M_{halo} \sim14.1$ (intermediate massive halos) and $\log M_{halo}\sim15.1$ (most massive halos).

The two models always reach similar conclusions.

In the mass range \M=9.4-10.5, galaxies in all the environments share a very similar mass distributions. At higher stellar masses, differences emerge, suggesting an effect of the environment only  for \M$>10.5$.
Very massive galaxies are hosted only in the most massive halos. As a consequence, the mass distribution characterizing the most massive halos is  flatter than those describing galaxies in the other environments. 
Comparing mass functions at different redshifts, simulations  find an evolution only in the massive end, in all environments. Both the shape and the extension of the mass functions change: in the local universe galaxies can reach higher stellar masses  and the number density of massive galaxies is higher than in the distant one. 

The dependence of the high mass end of the mass function on environment and its evolution in the high mass end are at odds with the observational results (e.g., C13, V13, \citealt{vanderburg13, ilbert10, pozzetti10, baldry12}).  
If both they are real, the discrepancies with the observational results might suggest that in observations uncertainties at high masses are still too large to firmly detect the simulated results. 

However, mass functions obtained from the DLB07 and G11 models are always different. 
In general, at \M$\leq10$, the G11 mass functions are flatter than the DLB07 ones, indicating that in the G11 prescription low mass galaxies are better treated and their excess is reduced. We remind that the 
 G11 model is an update of the DLB07 one, and differs from it  mainly for a different treatment of satellite evolution and a more efficient stellar feedback, as presented in \S\ref{theor_pred} (see also G11). 
These implementations reduce the tensions with observations, but there are still  residuals, indicating that the truncation mechanisms that influence the evolution of central and satellite galaxies still have to be better implemented.

\section{Summary and Conclusions}
In this paper we exploited two semi-analytic models (DLB07 and G11) to
 carefully investigate mass functions at different redshifts and in different environments, defined by the halo masses.  We have compared the theoretical findings of the DLB07 model with the observational results   published in \cite{drory09,morph}; C13 and V13. 
 
 The main results can be summarized as follows.

\begin{itemize}

\item Being able to match the field mass function (and its finer environments)  at $z=0$, the DLB07 model fails to reproduce the observed mass function of clusters at low-$z$ and over-predicts the number of low mass galaxies in both clusters and field at $z\sim0.6$. 

\item In sim-projections, the observed invariance of the mass function with the environment is reproduced for galaxies more massive than \M=10.25, the mass limit probed by the observations. On the other hand, the observed evolution of the shape of the mass function is not reproduced, neither in the filed  nor in clusters. 

\item Sim-projections reproduce well the observed stellar mass -  velocity dispersion relation, both for central galaxies and for the second most massive galaxies. The relation can be ascribed to the environment for the central galaxies at both redshifts and for the second most massive galaxies at low-$z$. On the other hand, it might be simply a statistical effect for  the second most massive galaxies at higher redshift. 

\item Inspecting only simulations, our results show that in both models the mass function  depends on the mass of the halo, both at $z=0.06 $ and $z=0.62$. In very massive halos there are proportionally more massive galaxies, with the result that the mass function is flatter and $M^*$ shifted toward higher values compared to lower mass halos. Simulations also detect a mass segregation with the environment: low-mass halos do not host massive galaxies.  

\item In both models, the overall shape of the mass function does not strongly depend on the halo centric distance, once  redshift and halo mass are controlled. However, subtle differences might be found when carefully inspecting the least massive systems and the high mass end of the most massive halos.

\item In both models, the shape of the mass function for \M$<$11.2 does not  evolve, in any environment. In contrast, there is an evolution in the number of most massive galaxies, which are more numerous in the local universe.  
\end{itemize}

The fact that at low-$z$ simulations fail
to reproduce the cluster mass function while they are successful in groups,
binary systems and isolated galaxies suggests
an incorrect inclusion of environmental processes in the models, such as tidal disruption, ram-pressure, strangulation. However, the disagreement between observations and simulations in all
environments at higher redshifts reveals that cluster physical processes
are not the only problem in the models and the 
redshift dependent evolution still needs to be better modeled. 

\begin{acknowledgements}
We thank the referee whose comments improved our manuscript.
This work was supported by World Premier International Research Center Initiative (WPI), MEXT, Japan.
The Millennium Simulation databases used in this paper and the web application providing online access to them were constructed as part of the activities of the German Astrophysical Virtual Observatory (GAVO).
GDL acknowledges financial support from the European Research Council under the European Community's Seventh Framework Programme (FP7/2007-2013)/ERC grant agreement n. 202781.
We are grateful to Ulisse Munari for providing us the conversions between different photometric systems.
We thank Niv Drory for providing  COSMOS stellar masses and the field mass functions, and  Fabio Fontanot 
for useful discussions.
\end{acknowledgements}

\appendix
\section{Consistency checks on  stellar masses}
\subsection{\cite{drory09} stellar masses}\label{drory_mass}

\cite{drory09} derived stellar masses  comparing multi-band photometry to a grid of stellar population models of varying star formation histories (SFH), ages, and dust content (for further details, refer to \citealt{drory09}). They adopted a \cite{chabrier03} IMF (0.1-100 $M_\odot$). 

\begin{figure}
\centering
\includegraphics[scale=0.55]{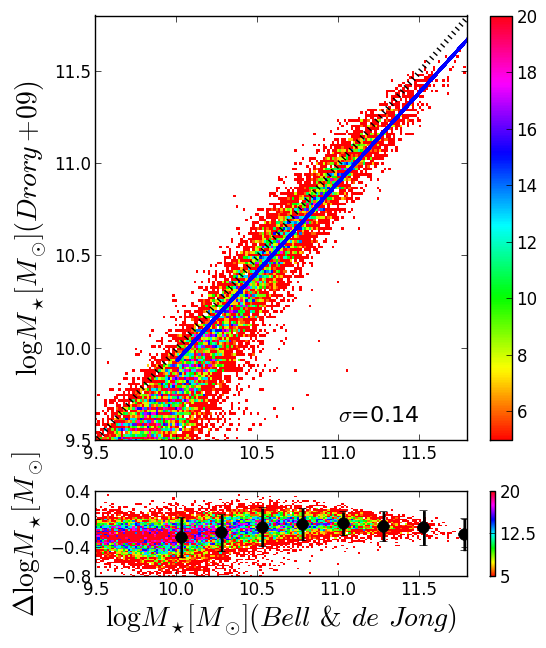}
\caption{Density maps showing the comparison between  stellar masses obtained using \cite{bj01} and those computed in \cite{drory09} and used to characterize the field mass function at $z=0.6$. Blue line represents the least square fit for galaxies with \M$>9.9$, the observed mass limit.  Black dotted line represents the 1:1 line.
The lower panel shows the difference between these two estimates. 
\label{cfr_masse_drory}}
\end{figure}

To reduce the uncertainties due to the different method adopted by \cite{drory09}, we computed stellar masses for the COSMOS sample using \cite{bj01} and then compared them to the stellar masses (Drory, private communication) used by \cite{drory09}. 
The comparison of the different mass estimates is presented in Fig.\ref{cfr_masse_drory}.
In addition to a large scatter, \cite{drory09} mass estimates are systematically smaller. Discrepancies are around $\sim0.2\, dex$, that corresponds to the typical error on the mass estimates. 
In our analysis, we apply a mean correction to the \cite{drory09} mass function  as a function of stellar mass,  taking into account the median difference between the two methods.

\subsection{DLB07 stellar masses}\label{model_mass}
\begin{figure*}
\centering
\includegraphics[scale=0.44]{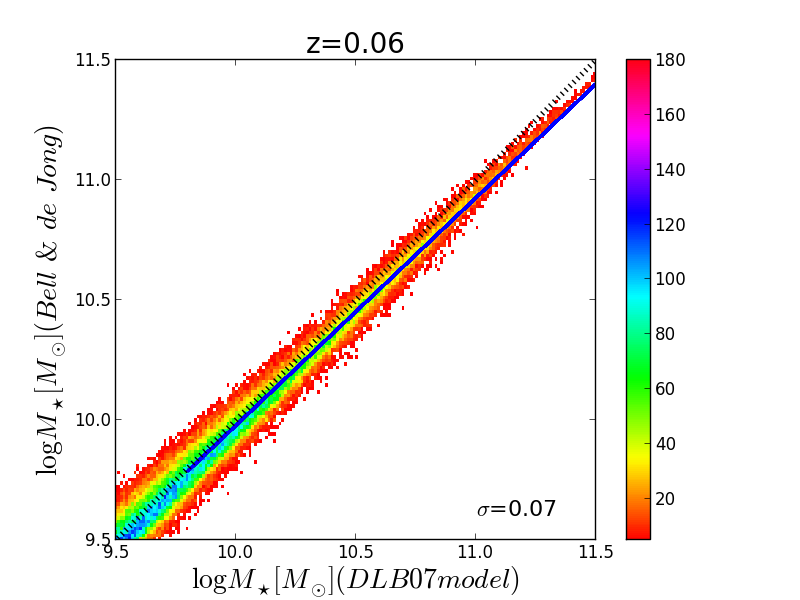}
\includegraphics[scale=0.44]{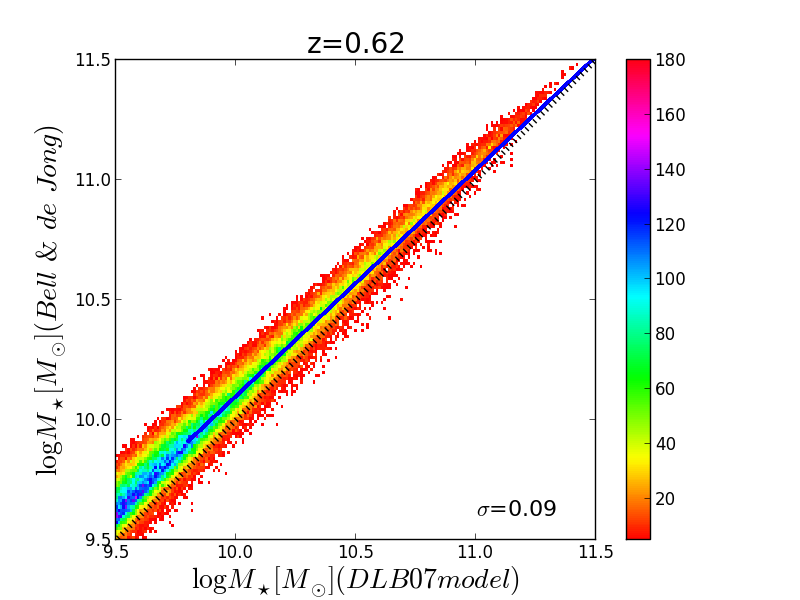}
\caption{Density maps showing the comparison between  stellar masses obtained using \cite{bj01} and those given by the DLB07 semi-analytic model for galaxies in all the samples. Left panel: low redshift; right panel: high redshift. Blue line represents the least square fit for galaxies with \M$>9.8$, our lowest observed mass limit.  Black dotted line represents the 1:1 line. 
\label{cfr_masse}}
\end{figure*}

In our analysis, when  semi-analytic model data are compared to observations, we adopt stellar masses computed 
using the \cite{bj01} formulation (Eq.\ref{bj}), as done for the observations.

Figure \ref{cfr_masse} shows the comparison between the two mass estimates, both obtained assuming  a \cite{kr01} IMF. 
Above \M=9.8 (our lowest observed mass limit), the estimates are 
in agreement within the typical scatter of the mass uncertainties  ($\sim 0.2-0.3 \, dex$,  see, e.g., \citealt{kannappan07}). The median difference of the masses 
is $\sim -0.08$ at  $z=0.06$, and  $\sim +0.05$  at $z=0.62$.

\section{The normalization of the mass functions}\label{app_normalization}

To compare mass functions in different environments, it is important to adopt normalizations which will allow a meaningful comparison. This is not always straightforward, since the number densities in clusters and in the field are expected to be very different. In numerical simulations of collisionless matter, cluster halos are defined as spherical regions enclosing a density which is 200 times the critical density ${\rho}_c(z)$ of the universe. Therefore the density of matter measured within cluster sized halos will be much larger than the average matter density. If galaxies are assumed to follow the dark matter distribution, then clusters should also have the same overdensity of galaxies within their boundaries with respect to the density of galaxies in the field. Our aim here is to remove this principal normalization difference to compare galaxy stellar mass functions in different environments. There could be additional corrections if galaxies do not exactly follow the dark matter distribution (e.g., if the number density profile  around a cluster does not follow a NFW profile or shows luminosity segregation) which we ignore for simplicity.

The volume normalization adopted for the field allows us to express the counts per unit comoving volume. To obtain the normalization for the cluster counts, note that ${\rho}_c(z)$ is related to the matter density $\bar{\rho}(z)$ as  
\begin{equation}\label{eq:rhoc_rhom}
{\rho}_c(z) =\bar{\rho}(z) \times \left [\Omega_0(1+z)^3+\Omega_\Lambda \right ]/\Omega_0(1+z)^3\,,
\end{equation}
where $\Omega_0$ is the matter density parameter and $\Omega_\Lambda$ is the dark energy density parameter and we have assumed a flat Universe. The number density of galaxies within $R_{200}$, should therefore be diluted by a factor
\begin{equation}
200 \times\left [\Omega_0(1+z)^3+\Omega_\Lambda \right ]/\Omega_0(1+z)^3\,
\end{equation}
while comparing to the number density in the field.

When we consider smaller regions in clusters (e.g.  $r<0.6R_{200}$),
we have to also take into account the density  variation as a function of the radius, as described by a \cite{nfw} profile. The mass enclosed within a radius  $\mathcal{X}=r/r_s$ is
\begin{equation}
M(<\mathcal{X})=M_{200}\times\frac{\mu(\mathcal{X})}{\mu(c_{200})}=\frac{4\pi}{3}R_{200}^3 200 \rho_c(z)\times\frac{\mu(\mathcal{X})}{\mu(c_{200})}
\end{equation}
where $c_{200}=R_{200}/r_s$ is the concentration parameter, $r_s$ a scale factor, $\mu(\mathcal{X})=\ln(1+\mathcal{X})-\mathcal{X}/(1+\mathcal{X})$. For relaxed halos, $c_{200}$ mildly  depends on the halo mass (see, e.g., \citealt{maccio08}). However, for our purpose, we can assume $c_{200}=5$.\footnote{In our halo mass range $c_{200}$ varies from 4 to 5.5 and this does not change our results.}

The density of matter within a radius $\mathcal{X}$ is
\begin{equation}
\rho(<\mathcal{X})=\frac{3\,M(<\mathcal{X})}{4\pi (r_s\mathcal{X})^3}=c_{200}^3\times 200{\rho}_c(z)\times\frac{\mu(\mathcal{X})}{\mu(c_{200})}.
\end{equation}
Therefore, using Equation~\ref{eq:rhoc_rhom}, the number counts within radius $\mathcal{X}$ should be scaled by the factor 
\begin{equation}
200 \times c_{200}^3 \times\frac{\mu(\mathcal{X})}{\mu(c_{200})} \times \left [\Omega_0(1+z)^3+\Omega_\Lambda \right ]/\Omega_0(1+z)^3.
\end{equation}
to compare with the number density in the field.

Similarly, when we consider a shell delimited by  $\mathcal{X}_2, \mathcal{X}_1$ (e.g. $1R_{200}<r<3R_{200}$), the appropriate normalization factor to use is 
\begin{equation}
200 \times \frac{c_{200}^3}{\mathcal{X}_2^3-\mathcal{X}_1^3} \times\frac{\mu(\mathcal{X}_2)-\mu(\mathcal{X}_1)}{\mu(c_{200})} \times \left [\Omega_0(1+z)^3+\Omega_\Lambda \right ]/\Omega_0(1+z)^3.
\end{equation}

\bibliographystyle{apj}

\bibliography{biblio}

\end{document}